\documentclass[12pt]{iopart}
\usepackage{iopams}
\usepackage{epsfig}
\eqnobysec
\newcommand{\AIII}{A{\rm I\!I\!I}}
\begin{document}
\title{Particle-Hole Symmetries in Condensed Matter}
\author{Martin R.\ Zirnbauer}
\address{Institut f\"ur Theoretische Physik, Universit\"at zu
K\"oln,\\ Z\"ulpicher Stra{\ss}e 77a, 50937 K\"oln, Germany}
\begin{abstract}
The term ``particle-hole symmetry'' is beset with conflicting meanings in contemporary physics. Conceived and written from a condensed-matter standpoint, the present paper aims to clarify and sharpen the termino\-logy. In that vein, we propose to define the operation of ``particle-hole conjugation'' as the tautological algebra auto\-morphism that simply swaps single-fermion creation and annihilation operators, and we construct its invariant lift to the Fock space. Particle-hole symmetries then arise for gapful or gapless free-fermion systems at half filling, as the concatenation of particle-hole conjugation with one or another involution that reverses the sign of the first-quantized Hamiltonian. We illustrate that construction principle with a series of examples including the Su-Schrieffer-Heeger model and the Kitaev-Majorana chain. For an enhanced perspective, we contrast particle-hole symmetries with the charge-conjugation symmetry of relativistic Dirac fermions. We go on to present two major applications in the realm of interacting electrons. For one, we argue that the celebrated Haldane phase of antiferromagnetic quantum spin chains is adiabatically connected to a free-fermion topological phase protected by a particle-hole symmetry. For another, we review the recent proposal by Son for a particle-hole conjugation symmetric effective field theory of the half-filled lowest Landau level, and we comment on the emerging microscopic picture of the composite fermion.
\end{abstract}
%
%\pacs{02.30.Fn, 02.50.Cw, 03.65.Nk, 05.30.Ch}
%
\vspace{2pc} \noindent{\it Keywords}: Tenfold Way classification of disordered fermions, particle-hole symmetries, charge conjugation, symmetry-protected topological phases, topological insulators and superconductors, Su-Schrieffer-Heeger model, Kitaev-Majorana chain, Hubbard model at half filling, antiferromagnetic quantum spin chains, Haldane phase, half-filled lowest Landau level, composite fermions

\maketitle

\tableofcontents

\section{Introduction}

"Particle-hole symmetry" is a term frequently encountered in contemporary physics, yet it seems to have no canonical meaning. It has been the author's long-standing tenet \cite{HHZ,OHRMT,KZ-PS,AMZ} that particle-hole symmetries for electrons in condensed matter are to be defined as complex \emph{antilinear} operations that \emph{commute} with the quantum Hamiltonian. However, differing usages (such as: antilinear and anti-commute with the Hamiltonian; linear and commute with the Hamiltonian; etc.) abound. Perhaps most disturbingly, sometimes one is confronted with the puzzling statement that all superconductors are particle-hole symmetric! If that was really so, wouldn't one have to concede to a sceptic that particle-hole symmetry is a redundant or even meaningless concept?

Partly review and partly original material, the present paper grew out of a colloquium talk \cite{MRZ-2016} and an oversized chapter in a planned article \cite{MRZ-LF} surveying the Tenfold Way of symmetry classes for disordered fermions \cite{HHZ} --- it is the author's attempt to offer clarification of the multi-faceted and confusing topic in its title. On the theory side, we will define particle-hole symmetry \emph{invariantly}, i.e., without fixing any preferred single-particle basis of Hilbert space; and we propose to sharpen the terminology by making a distinction between particle-hole symmetry transformations and the tautological operation of particle-hole conjugation. On the applied side, we will illustrate the theoretical concept with numerous examples from condensed matter physics, including the Su-Schrieffer-Heeger model, Kitaev's Majorana chain, the Hubbard model at half filling, the antiferromagnetic Heisenberg quantum spin chain, and the fascinating story of composite fermions in the half-filled lowest Landau level.

To put our notion of particle-hole symmetry into context, let us begin by visiting a close cousin: charge-conjugation symmetry, $C$, for relativistic Dirac fermions. In the precursory framework of relativistic quantum mechanics (i.e., for the Dirac equation as a first-quantized theory) charge conjugation is complex antilinear, which means that it sends complex scalars to their complex conjugates \cite{BjorkenDrell,Thaller}. Perhaps confusingly to the novice, the antilinear property transmutes to linear (!) when one passes (by second quantization) to the final interpretation of the Dirac equation as a quantum field theory with electron and positron excitations of positive energy with respect to the Dirac sea of filled negative-energy states. In fact, charge conjugation $C$ of the Dirac quantum field is a unitary (hence complex linear) symmetry exchanging electrons with positrons \cite{Weinberg}.

Why is charge conjugation complex antilinear in first quantization but complex linear in second quantization? The reason is quite simple. Let the Hilbert space $V$ of the first-quantized Dirac theory be decomposed into the spaces $V_{\pm}$ of positive- and negative-energy states, and let $V_{\pm}^\ast$ be their duals. The fermionic Fock space $\mathcal{F}$ of the quantum field theory is an exterior algebra $\mathcal{F} = \bigwedge(V_+) \otimes \bigwedge(V_-^\ast)$, generated from the true vacuum (the filled Dirac sea) by creating particles in $V_+$ and holes in $V_-$ (by exterior multiplication with co-vectors from $V_-^\ast$). Now charge conjugation initially is an exchange mapping $V_+ \leftrightarrow V_-$ and $V_-^\ast \leftrightarrow V_+^\ast\,$, which gives rise to a canonical (yet unphysical) correspondence $\bigwedge(V_+) \otimes \bigwedge(V_-^\ast) \leftrightarrow \bigwedge(V_-) \otimes \bigwedge(V_+^\ast)$. To arrive at a physically meaningful operation taking $\mathcal{F}$ to itself, one has no choice but to make some identification of Hilbert vectors with co-vectors. In the absence of extraneous information, there exists only one such identification: the Dirac ket-to-bra bijection (or Fr\'echet-Riesz isomorphism) $V \to V^\ast$, $\vert v \rangle \mapsto \langle v \vert\,$. Thus the finished product $C :\; \mathcal{F} \to \mathcal{F}$ may be viewed as a concatenation of two steps: first, $\mathcal{F}$ is mapped to $\bigwedge (V_+^\ast) \otimes \bigwedge(V_-)$ by the charge-conjugation map of first quantization, which is antilinear; and in the second step one returns to the physical space $\mathcal{F}$ by means of the bijection $\vert v \rangle \leftrightarrow \langle v \vert\,$, which is also antilinear. Since two complex conjugations amount to nothing, $C$ ends up being complex linear (or linear, for short).

After this overture, we move on to our subject proper: particle-hole symmetries in condensed-matter systems. The author's interest in the subject originated from the so-called ``Tenfold Way'' of symmetry classes for disordered free fermions \cite{AZ97,HHZ}, which we now summarize very briefly for its historical relevance. The Tenfold Way extends the ``Threefold Way'' of Dyson \cite{dyson}, a classification scheme for random-matrix ensembles by symmetries. Dyson's scheme rests on the mathematical setting of a Hilbert space carrying the action of a symmetry group generated by any number of unitary operations together with one distinguished anti-unitary, namely time reversal. Our extended setting in \cite{HHZ} follows Dyson in that it builds on the \underline{same} fundamental notion of symmetry \cite{footnote0}: symmetries always \emph{commute} with the Hamiltonian, never do they anti-commute! The main novelty in the Tenfold Way setting as compared with Dyson's is that plain Hilbert space is replaced by the more refined structure of a Fock space for fermions (thereby inviting generalizations to systems of interacting particles). Very importantly, the Fock space setting makes for the additional option of a particle-hole transformation as part of the symmetry group. In that extended setting of a fermionic Fock space, and in the possible presence of a particle-hole symmetry, the author, with Heinzner and Huckleberry, established a 10-way classification scheme \cite{HHZ} of Hamiltonians $H$ for ``free fermions'', meaning interacting and/or disordered fermions treated in the most general mean-field approximation, which is known by the names of Hartree, Fock, and Bogoliubov. That scheme had been anticipated on heuristic grounds in \cite{AZ97}; its recent applications will be reviewed from a modern perspective in a companion article \cite{MRZ-LF}.

As was stated at the outset, by a ``particle-hole symmetry'' we mean an \emph{antilinear} operation that commutes with the Hamiltonian. (Thus, particle-hole symmetries differ from charge-conjugation symmetry, which is linear.) Here it must be emphasized that a particle-hole symmetry, as a true physical symmetry, leaves the many-body ground state invariant and turns excitations of energy $E$, electric charge $Q$, and electric current $I$, into excitations of the \emph{same} energy $+E$ and current $+I$ but the \emph{opposite} charge $-Q$. (In contrast, the time-reversal operator leaves the energy and charge the same, but reverses the current.) Please be warned that much of the recent literature (ab-)uses the word ``particle-hole symmetry'' for an algebraic tautology that derives from nothing but the canonical anti-commutation relations for fermion fields! Our proposal is to reserve the word \emph{particle-hole conjugation}, with symbol $\Xi$, for that very operation. Note that $\Xi$ simply exchanges single-particle creation and annihilation operators ($a^\dagger \leftrightarrow a$) and can never be a symmetry for free fermions. In fact, any self-adjoint and traceless one-body Hamiltonian $H$ is \emph{odd} under particle-hole conjugation ($\Xi H \Xi^{-1} = - H$).

In preparing this article, the author went back and forth on the question as to which math symbol to adopt for particle-hole symmetries. In the original reference \cite{HHZ}, particle-hole symmetries were called ``mixing symmetries'' (mixing particles with holes) and denoted by $T$ or $T_1$. Needless to say, that nomenclature did not stand the test of time. In a previous short review \cite{OHRMT}, the author changed the language to ``twisted particle-hole conjugation'', $\tilde{C}$. Alas, that choice was also ill-conceived, as the symbol $C$ is canonical for charge conjugation. In the present article, we will use the letter $K$ \cite{footnote}.

To provide a little more detail and sharpen our motivation, let us mention here that particle-hole symmetries $K$ are relevant for condensed-matter electrons at ``half filling'' (akin to the relativistic setting with a filled Dirac sea): given a decomposition of the single-particle Hilbert space $V = V_+ \oplus V_-$ by two subspaces $V_\pm$ (of single-electron energy above resp.\ below the Fermi level), the ground state of the free-electron system is the filled Fermi sea $\bigwedge^{\rm top}(V_-)$, and the Fock space $\bigwedge(V_+ \oplus V_-^\ast)$ of the many-body system is then built from particle-like excitations in $\bigwedge(V_+)$ and hole-like excitations in $\bigwedge(V_-^\ast)$. A particle-hole symmetry $K$ is usually an antilinear transformation that exchanges the two building blocks $\bigwedge(V_+)$ and $\bigwedge(V_-^\ast)$ of the Fock space (assuming the two vector spaces $V_+$ and $V_-$ to be isomorphic). Examples of such a situation, foremost the Su-Schrieffer-Heeger model \cite{SSH}, are expounded in detail in the body of the paper.

The contents are as follows. We start out in Section \ref{sect:gapped} by defining the notion of a particle-hole symmetry $K$ for gapped systems such as band insulators, where the chemical potential resides in an energy gap separating the conduction and valence bands. We illustrate the definition with three examples: the Su-Schrieffer-Heeger model (SSH), the Kitaev-Majorana chain, and relativistic Dirac fermions. We also explain how systems with chiral ``symmetry'' fit into the picture, and we briefly describe the manifestation of particle-hole symmetries in the real-time path integral. In Section \ref{sect:gapless}, we take up the more challenging case of gapless systems. Focusing on the subspace of zero modes (or close-to-zero modes), we introduce the tautological operation of particle-hole conjugation $\Xi$ first by its action on the operators and then construct its lift to the Fock space. Particle-hole symmetries for gapless free fermions are obtained as $K = \Xi \circ \Gamma$, by concatenating $\Xi$ with an involution $\Gamma$ that inverts the sign of the first-quantized Hamiltonian. In Section \ref{sect:interact} we turn to interacting systems. We review the particle-hole symmetry of the Hubbard model and the antiferromagnetic Heisenberg quantum spin chain. We also discuss how symmetry-protected topological phases of interacting fermions in 1D with symmetry group $\mathrm{U}(1) \times \mathbb{Z}_2^K$ (a.k.a.\ type $\AIII$) are classified by their gapless end modes. Section \ref{sect:FF2Haldane} presents the first of two major applications. Starting from a free-fermion Hamiltonian for two spinful SSH chains, we deform to the $S = 1$ antiferromagnetic Heisenberg chain by turning on a repulsive Hubbard interaction and an attractive Hund's rule coupling. In this way we argue that the celebrated Haldane phase is adiabatically connected to a free-fermion topological phase protected by a particle-hole symmetry. Our second major application, in Section \ref{sect:LLL}, is the recent proposal \cite{DTSon} for a particle-hole (conjugation) symmetric effective field theory of the half-filled lowest Landau level. We review its striking symmetry aspects in first and second quantization, and we comment on the emerging microscopic picture for the composite fermion. We conclude with a summary of our key messages in Section \ref{sect:conclude}.

\section{Gapped systems (band insulators)}\label{sect:gapped}

To organize our exposition, we shall distinguish between systems with and without an energy gap for excitations of the ground state. We begin with the former kind.

\begin{figure}
    \begin{center}
        \epsfig{file=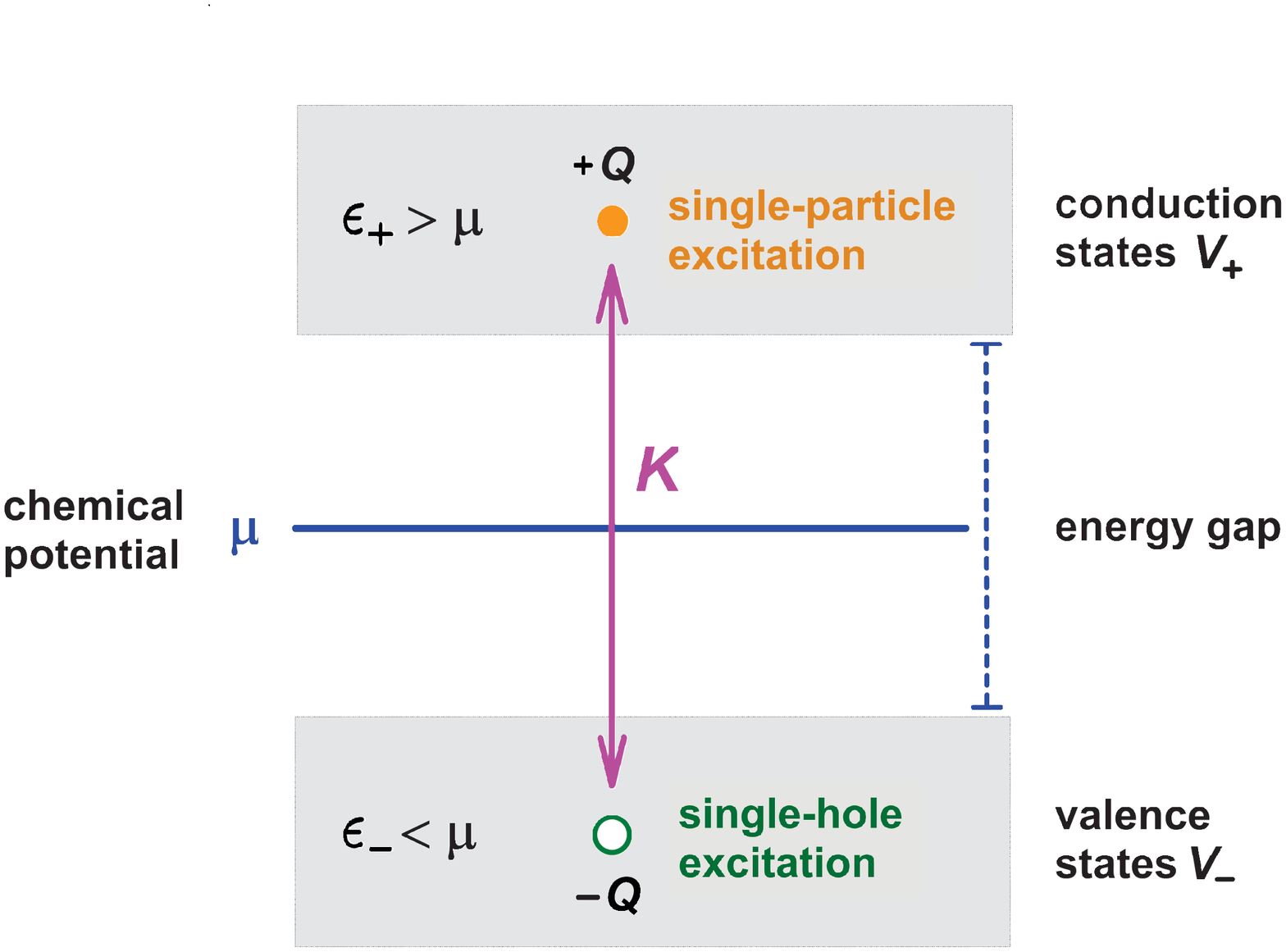,height=8cm}
    \caption{Particle-hole symmetry of a band insulator. The chemical potential $\mu$ lies in an energy gap separating the conduction ($V_+$) from the valence ($V_-$) states. The latter/former are filled/empty in the free-fermion ground state. A particle-hole symmetry $K$ exchanges single-particle excitations in $V_+$ of energy $E = \epsilon_+ - \mu > 0$ and charge $Q$ with single-hole excitations in $V_-$ of the same energy $E = \mu - \epsilon_-$ and the opposite charge $-Q$.} \label{fig:withgap}
    \end{center}
\end{figure}

The defining property of a band insulator is that the chemical potential $\mu$ (a.k.a.\ Fermi level) lies in an energy gap of the band structure or the single-particle energy spectrum (Fig.\ \ref{fig:withgap}). That gap separates the single-particle Hilbert space $V$ into the subspace $V_+$ of conduction states with single-particle energy above the gap and the subspace $V_-$ of valence states with s.p.\ energy below the gap. (Note that here we do \underline{not} assume $V_+$ or $V_-$ to have finite dimension. In fact, both may be infinite-dimensional.)

The formalism of many-body quantum theory for a band insulator or similar  gapped system is built on the free-fermion ground state obtained by completely filling the valence subspace. The elementary excitations of that ``Fermi sea'' are created by adding a single particle in the conduction subspace $V_+$ or adding a hole (i.e., removing a particle) in the filled valence subspace $V_-\,$. Mathematically speaking,
the subspace of particle excitations can be identified with the exterior algebra ${\bigwedge}(V_+)$, and the subspace of hole excitations is $\bigwedge(V_-^\ast)$ with $V_-^\ast$ the vector space dual to $V_-$. The total Fock(-Hilbert) space $\mathcal{F}$ generated by all such excitations is the exterior algebra $\bigwedge(V_+) \otimes \bigwedge(V_-^\ast) \cong \bigwedge(V_+ \oplus V_-^\ast)$ with Hermitian scalar product induced by that of the single-particle Hilbert space $V = V_+ \oplus V_-\,$. Our so-defined Fock space is bi-graded:
\begin{equation*}
    \mathcal{F} \equiv {\textstyle{\bigwedge}}(V_+ \oplus V_-^\ast) = \bigoplus_{p,q \geq 0} {\textstyle{\bigwedge}}^{p,q}(V_+ \oplus V_-^\ast) ,
\end{equation*}
with summands given by
\begin{equation*}
    {\textstyle{\bigwedge}}^{p,q}(V_+ \oplus V_-^\ast) \cong {\textstyle{\bigwedge}}^p (V_+) \otimes {\textstyle{\bigwedge}}^q (V_-^\ast) ,
\end{equation*}
where $p$ (resp.\ $q$) is the number of single-particle (resp.\ single-hole) excitations. The difference $p-q$ is called the charge of the excited state. We write $\bigwedge^{p,q}(V_+ \oplus V_-^\ast) \equiv \mathcal{F}^{p,q}$ for short. The subspace $\mathcal{F}^{0,0} \cong \mathbb{C}$ is the complex line of the Fermi sea (or Fock vacuum).

\smallskip\noindent\emph{Definition.} By a \emph{particle-hole transformation} $K$ we mean a charge-reversing map
\begin{equation}\label{eq:Def-C}
    K : \; \mathcal{F}^{p,q} \to \mathcal{F}^{q,p} .
\end{equation}
$K$ is called a symmetry of the second-quantized Hamiltonian $H$ if $K H = H K$. $\square$ \smallskip

In the following, we will look at particle-hole transformations in more detail. To uncover their true nature and offer the best possible perspective, we will make an effort to define all our operations in invariant terms. Thus we use the basis-free notation
\begin{equation*}
    \varepsilon(v_+) :\; \mathcal{F}^{p,q} \to \mathcal{F}^{p+1,q} , \quad \varepsilon(\varphi_-) :\; \mathcal{F}^{p,q} \to \mathcal{F}^{p,q+1},
\end{equation*}
with $v_+ \in V_+$ and $\varphi_- \in V_-^\ast$ for the Fock operators that increase the number of elementary excitations of the Fermi sea and
\begin{equation*}
    \iota(\varphi_+) :\; \mathcal{F}^{p,q} \to \mathcal{F}^{p-1,q} , \quad \iota(v_-) :\; \mathcal{F}^{p,q} \to \mathcal{F}^{p,q-1} ,
\end{equation*}
with $\varphi_+ \in V_+^\ast$ and $v_- \in V_-$ for the Fock operators annihilating elementary excitations. The symbol $\varepsilon(\bullet) \equiv \bullet \, \wedge$ means the wedge or exterior product with vectors $v_+ \in V_+$ or co-vectors $\varphi_- \in V_-^\ast$, and $\iota(\bullet)$ stands for the inner product or alternating contraction with co-vectors $\varphi_+ \in V_+^\ast$ or vectors $v_- \in V_-$. These creation and annihilation operators satisfy the canonical anti-commutation relations for fermions. Thus they generate a Clifford algebra of many-body operators acting on the fermionic Fock space $\mathcal{F}$. The non-vanishing anti-commutators are
\begin{eqnarray*}
    &&\iota(\varphi_+) \varepsilon(v_+) + \varepsilon(v_+) \iota(\varphi_+) = \langle \varphi_+ , v_+ \rangle \, \mathbf{1}_{\mathcal{F}}  \,, \cr &&\varepsilon(\varphi_-) \iota(v_-) + \iota(v_-) \varepsilon(\varphi_-) = \langle \varphi_- , v_- \rangle \, \mathbf{1}_{\mathcal{F}}  \,,
\end{eqnarray*}
where $\langle \varphi , v \rangle \equiv \varphi(v)$ means the canonical pairing between vector and co-vector.

The basis-free notation for Fock operators is converted to standard physics notation by choosing an orthonormal basis $\vert j \rangle$ of $V$ with $\vert j \rangle \in V_+$ for $j > 0$ and $\vert j \rangle \in V_-$ for $j < 0$. The dual basis of $V^\ast$ is denoted by $\langle j \vert$. One then writes
\begin{equation*}
    \varepsilon( \vert j \rangle ) \equiv a_j^\dagger \,, \quad
    \iota( \langle j \vert ) \equiv a_j \,, \quad
    \iota( \vert l \rangle ) \equiv a_l^\dagger \,, \quad
    \varepsilon( \langle l \vert ) \equiv a_l \qquad (j > 0 > l).
\end{equation*}
The Fock vacuum $\mathcal{F}^{0,0} \cong \mathbb{C}$ with state vector $\vert {\rm vac} \rangle$ is determined by
\begin{equation*}
    a_j |{\rm vac} \rangle = 0 = a_l^\dagger |{\rm vac}\rangle \quad (j > 0 > l) .
\end{equation*}

For completeness and later reference, we are going to write down how the Hamiltonian $h$ of a band insulator in first quantization acts as a second-quantized Hamiltonian $H$ on the Fock space $\bigwedge(V_+ \otimes V_-^\ast)$. Defined as an operator on $V = V_+ \oplus V_-$, the first-quantized Hamiltonian decomposes into blocks,
\begin{equation}\label{eq:1QMH}
    h = \Bigg( \begin{array}{cc} A &B \\ C &D \end{array} \Bigg) \in
    \Bigg( \begin{array}{cc} \mathrm{End}(V_+) &\mathrm{Hom}(V_-,V_+) \\ \mathrm{Hom}(V_+,V_-) &\mathrm{End}(V_-) \end{array} \Bigg) .
\end{equation}
Then, taking the chemical potential $\mu$ to vanish (by the choice of zero on the energy axis) and using the basis for $V = V_+ \oplus V_-$ from above, the normal-ordered second-quantized Hamiltonian is expressed as
\begin{eqnarray}\label{eq:2QMH}
    H &=& \sum_{j>0} \varepsilon(A | j \rangle) \iota(\langle j |)
    + \sum \varepsilon(B |l\rangle) \varepsilon(\langle l |) \cr
    &-& \sum \iota(\langle j |) \iota(C | j \rangle)
    - \sum_{l<0} \varepsilon(\langle l |) \iota(D | l \rangle ) .
\end{eqnarray}
Here the terms $\varepsilon \, \varepsilon :\; \mathcal{F}^{p,q} \to \mathcal{F}^{p+1,q+1}$ create particle-hole excitations (without changing the charge $p-q$ of the state), while their Hermitian conjugates $\iota\,\iota :\; \mathcal{F}^{p,q} \to \mathcal{F}^{p-1,q-1}$ do the opposite.

\subsection{Particle-hole transformation as a concatenation}
\label{sect:3.2.1}

To provide a constructive view of the operator $K$ of a particle-hole transformation, we introduce two mappings. Firstly, let $\Gamma$ be an involutive isomorphism $\Gamma :\; V \to V$ exchanging the conduction and valence subspaces ($\Gamma V_\pm = V_\mp\,$, $\Gamma^2 = \mathbf{1}$). Secondly, recall the definition of the Dirac ket-to-bra bijection $\vert j \rangle \mapsto \langle j \vert$ (or, mathematically speaking, the Fr\'echet-Riesz isomorphism) by
\begin{equation*}
     c :\; V \to V^\ast , \quad v \mapsto \langle v \vert \,\cdot\, \rangle_V \,.
\end{equation*}
Then by concatenating $c$ with $\Gamma$ we get a mapping
\begin{equation*}
    c \circ \Gamma :\; V \to V^\ast , \quad V_\pm \ni v_\pm \mapsto c(\Gamma v_\pm) = \langle \Gamma v_\pm \vert \,\cdot\, \rangle_V \in V_\mp^\ast \,,
\end{equation*}
sending $V_+$ to $V_-^\ast$ and $V_-$ to $V_+^\ast \,$. By second quantization, $c \circ \Gamma$ induces a mapping on Fock space:
\begin{eqnarray}
     K :\; &&v_{+1} \wedge \cdots \wedge v_{+p} \wedge \varphi_{-1} \wedge \cdots \wedge \varphi_{-q} \nonumber \\ \mapsto &&c(\Gamma v_{+1}) \wedge \cdots \wedge c(\Gamma v_{+p}) \wedge \Gamma c^{-1} (\varphi_{-1}) \wedge \cdots \wedge \Gamma c^{-1}(\varphi_{-q}) ,
    \label{eq:C-Fock}
\end{eqnarray}
which is a particle-hole transformation $K :\; \mathcal{F}^{p,q} \to \mathcal{F}^{q,p}$ of the type of (\ref{eq:Def-C}). We repeat that a system with Hamiltonian $H$ is called \emph{particle-hole symmetric} if $K H K^{-1} = H$.

{}From the definition (\ref{eq:C-Fock}) and the fact that the Hermitian adjoint of $\varepsilon(v)$ is $\iota(c(v))$, it follows immediately that the particle-hole transformation $K$ acts on the basic Fock operators as
\begin{eqnarray*}
    &&K \varepsilon(v_+) K^{-1} = \varepsilon(c(\Gamma v_+)) , \quad
     K \varepsilon(\varphi_-) K^{-1} = \varepsilon(\Gamma c^{-1}(\varphi_-)) , \\ && K \iota(v_-) K^{-1} = \iota(c(\Gamma v_-)) , \quad \, K \iota(\varphi_+) K^{-1} = \iota(\Gamma c^{-1}(\varphi_+)) .
\end{eqnarray*}
Thus creation operators particle-hole transform to creation operators ($\varepsilon \to \varepsilon$), and the same goes for the annihilation operators ($\iota \to \iota$).

We note that the Fr\'echet-Riesz isomorphism $c$ is antilinear by its definition from the Hermitian scalar product $\langle \cdot \vert \cdot \rangle_V$. While $\Gamma$ may, in principle, be linear or antilinear, it is linear in all the examples we know from condensed matter physics. Thus, for all our CMP purposes, the particle-hole transformation $K$ will be antilinear.

Assuming $\Gamma$ to be linear and self-adjoint ($\Gamma^\dagger = \Gamma$), we obtain the particle-hole transform of the second-quantized Hamiltonian (\ref{eq:2QMH}) as
\begin{eqnarray}\label{eq:H-phK}
    K H K^{-1} = &-& \sum_{l<0} \varepsilon(\Gamma |l\rangle) \iota(\langle l | D \Gamma) - \sum \varepsilon(\Gamma | l \rangle) \varepsilon(\langle l | C \Gamma)  \cr
    &+& \sum \iota(\langle j | B \Gamma) \iota(\Gamma | j \rangle) + \sum_{j>0} \varepsilon(\langle j | A \Gamma) \iota(\Gamma | j \rangle) ,
\end{eqnarray}
where the relations $A^\dagger = A$, $D^\dagger = D$, and $B^\dagger = C$ (from $h^\dagger = h$) have been used. Before we elucidate what the condition of particle-hole symmetry $K H K^{-1} = H$ implies for the first-quantized Hamiltonian $h$, we shall visit a series of examples.

\subsection{Example 1: Su-Schrieffer-Heeger model}\label{sect:SSH}

One of the simplest examples of a particle-hole symmetric Hamiltonian is that of a cosine band for electrons with trans\-lation-invariant hopping between the nearest neighbors of a chain of atomic sites. The Hamiltonian is diagonal in the momentum representation:
\begin{equation}\label{eq:H-cosband}
    H = \int \frac{dk}{2\pi}\, t(k) :\! a_k^\dagger a_k^{\vphantom{ \dagger}} \!: \,, \quad t(k) = - t_0 \, \cos k , \quad t_0 > 0 ,
\end{equation}
where the standard physics notation for the Fock operators is used. Labeling the atomic sites by the integers, one has a conserved (quasi-)momentum in the range of $k \in \mathbb{R} / 2\pi\mathbb{Z}$. The colons around the operator mean normal ordering with the respect to the free-fermion ground state. The non-vanishing anti-commutator is $a_k^\dagger a_{k^\prime}^{\vphantom{\dagger}} + a_{k^\prime}^{ \vphantom{\dagger}} a_k^\dagger = 2\pi \delta(k - k^\prime)$.

\begin{figure}
    \begin{center}
        \epsfig{file=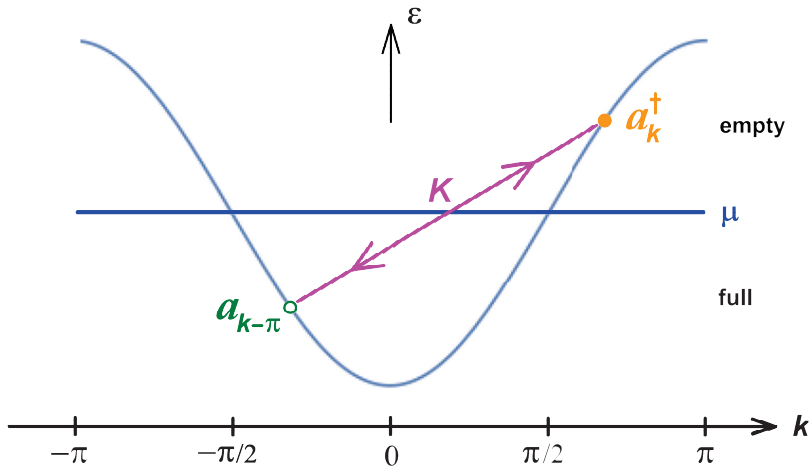,height=8cm}
    \caption{Particle-hole symmetry of the cosine band at half filling. The particle-hole transformation $K$ exchanges the single-particle creation operator at momentum $k$ with the single-hole creation operator at the shifted momentum $k\pm\pi$.}
    \label{fig:cosband}
    \end{center}
\end{figure}

To put our system at half filling, we place the Fermi level at zero energy, so that the valence subspace $V_- \subset V$ is spanned by the momentum eigenstates with $\vert k \vert < \pi/2$, where $t(k) < 0$, while the conduction space $V_+$ is spanned by the remaining states, where $t(k) > 0$. In that situation, the Hamiltonian (\ref{eq:H-cosband}) has a particle-hole symmetry by an operator $K :\; \mathcal{F} \to \mathcal{F}$ that acts on the Fock operators as
\begin{equation*}
     K a_k^\dagger K^{-1} = a_{k+\pi} \,, \quad K a_k K^{-1} = a_{k+\pi}^\dagger \,,
\end{equation*}
cf.\ Fig. \ref{fig:cosband}. The involutive isomorphism $\Gamma :\; V_\pm \to V_\mp$ is given here by the mapping that shifts the momentum $k \to k + \pi$. Note that $K$ is necessarily antilinear ($K \mathrm{i} K^{-1} = - \mathrm{i}$) as there exists {\it a priori} no basis-independent way to make such an exchange as $a_k \leftrightarrow a_k^\dagger$ without invoking the Dirac ket-to-bra bijection $\vert k \rangle \leftrightarrow \langle k \vert$, which is antilinear.

The $K$-symmetry of the second-quantized Hamiltonian (\ref{eq:H-cosband}) is easily verified:
\begin{equation*}
    K H K^{-1} = \int \frac{dk}{2\pi} \, \overline{t(k)} :\! a_{k+\pi}^{\vphantom{\dagger}} a_{k+\pi}^\dagger \!: \; = - \int \frac{dk}{2\pi}\, t(k-\pi) :\! a_k^\dagger a_k^{\vphantom{\dagger}} \!: \; = H ,
\end{equation*}
using that $t(k)$ is real, and $- t(k-\pi) = t(k)$. Note that no constant term is generated by the restoration of normal ordering, as the period integral $\int t(k) dk  = - \int t(k) dk$ vanishes. Note also that the particle-hole symmetry $K H K^{-1} = H$ holds more generally for any dispersion relation with the property $t(k \pm \pi) = - t(k)$.

Now the Hamiltonian (\ref{eq:H-cosband}) does not have an energy gap. To stay in line with our present framework for gapped systems, we may appeal to the Peierls instability and replace the translation-invariant hopping by staggered hopping, thereby reducing the group of translation symmetries from $\mathbb{Z}$ to $2\mathbb{Z}$. We accordingly double the unit cell in real space (and halve the Brillouin zone in momentum space). By what is known as ``back folding'' over the reduced Brillouin zone, we obtain a setting with one valence and one conduction band. The staggered hopping opens a gap between the two bands. Alternatively, we can stick with a single band, letting $t(k)$ jump through zero at $k = \pm \pi/2$ (mod $\pi$). Either way, we arrive at a model with particle-hole symmetry and an energy gap; that model goes by the names of Su, Schrieffer, and Heeger \cite{SSH}. We will consider some of its extensions (especially, to interacting systems) below.

\subsection{Example 2: Kitaev(-Majorana) chain}\label{sect:Kitaev}

In the first example addressed above, we looked only at the simple case of free fermions with conserved particle number. Yet the notion of particle-hole symmetry $K$ makes perfect sense for more general systems. Staying within the realm of free fermions, we may consider Hamiltonians of the Hartree-Fock-Bogoliubov (HFB) mean-field type:
\begin{equation}\label{eq:BdG}
    H = \sum h_{kk^\prime}^{\vphantom{\dagger}}\, a_k^\dagger a_{k^\prime}^{\vphantom{\dagger}} + {\textstyle{\frac{1}{2}}} \sum \left( \Delta_{kk^\prime}^{\vphantom{\dagger}}\, a_k^\dagger a_{k^\prime}^\dagger + \overline{\Delta}_{kk^\prime}\, a_{k^\prime} a_k \right) ,
\end{equation}
where the indices $k$, $k^\prime$ refer to an orthonormal basis of
the single-particle Hilbert space. A special example of such an HFB mean-field Hamiltonian is the Kitaev chain \cite{Kitaev-BDI}:
\begin{equation}\label{eq:kitaev}
    H = \frac{t}{2} \sum (a_{x+1}^{\vphantom{\dagger}} - a_{x+1}^\dagger) (a_x^{\vphantom{\dagger}} + a_x^\dagger) ,
\end{equation}
which models a superconductor of spinless fermions $a_x\,$, $a_x^\dagger$ moving on a chain of sites labeled by $x \in \mathbb{Z}$. As before, the many-particle system is put at half filling by taking the chemical potential to be zero.

The Kitaev chain is usually advertised as a system with anti-unitary time-reversal symmetry $T$ obeying $T^2 = +1$:
\begin{equation*}
    T a_x T^{-1} = a_x \,, \quad T a_x^\dagger T^{-1} = a_x^\dagger \,, \quad T \mathrm{i} T^{-1} = - \mathrm{i} \,.
\end{equation*}
While that perspective is mathematically viable, it begs the question as to how one is supposed to make physical sense of the fermions $a_x\,$. (In fact, for fundamental fermions one has $T^2 = (-1)^F$, and the effective relation $T^2 = +1$ for fermion number $F = 1$ emerges only after symmetry reduction for systems with $\mathrm{SU}(2)$ spin-rotation invariance.)

A different perspective on the Kitaev chain is that it has an anti-unitary particle-hole symmetry $K H K^{-1} = H$ with
\begin{equation}\label{eq:ph-Kitaev}
    K a_x K^{-1} = (-1)^x a_x^\dagger \,, \quad
    K a_x^\dagger K^{-1} = (-1)^x a_x\,,
\end{equation}
as is easily verified. [Incidentally, this coincides with the particle-hole transformation $K$ for the cosine band (\ref{eq:H-cosband}) transcribed to position space.] The ground state of the Kitaev chain is fully characterized by the complex lines $A_k$ of its quasi-particle annihilation operators. By Fourier transformation these are
\begin{equation*}
    A_k = \mathbb{C} \alpha_k \,,\quad \alpha_k = \sum\nolimits_x \mathrm{e}^{\mathrm{i} kx} \left( \mathrm{i} \sin(k/2)\, a_x + \cos(k/2)\, a_x^\dagger \right) .
\end{equation*}
Indeed, one easily checks that the $\alpha_k$ obey the commutation relation $[H , \alpha_k] = - t \alpha_k$ and thus lower the energy by $t > 0$. The state of lowest energy, known as the superconducting ground state in the HFB mean-field approximation, is the one annihilated by all these energy-lowering operators. Note that although $\alpha_k$ is double-valued as a function of $k \in \mathbb{R} / 2\pi \mathbb{Z}$, the line bundle $A_k = \mathbb{C} \alpha_k \mapsto k$ of annihilation spaces $A_k$ is well-defined.

By applying the particle-hole transformation (\ref{eq:ph-Kitaev}) to the annihilation operators $\alpha_k$ one finds that
\begin{equation}
    K A_k K^{-1} = A_{\pi - k} \,.
\end{equation}
Thus, quasi-particle annihilation operators are particle-hole transformed into other quasi-particle annihilation operators (and the same holds true for the quasi-particle creation operators). By consequence, the superconducting many-body ground state is particle-hole symmetric. In summary, what we are seeing here is another instance of a true particle-hole symmetry. The difference from before is that the ground state and its excitations do not carry definite charge.

\subsection{Chiral ``symmetry''}

As we have explained, a particle-hole transformation $K$ is constructed from two pieces: (i) an involutive isomorphism $\Gamma : V_\pm \to V_\mp$ of the single-particle Hilbert space $V$, and (ii) the Fr\'echet-Riesz antilinear isomorphism $c :\; V \to V^\ast$. The concatenation $c \circ \Gamma$ acts on the Fock space $\mathcal{F} = \bigwedge(V_+ \otimes V_-^\ast)$ as $K$ by (\ref{eq:C-Fock}). Defined in this way, the notion of particle-hole symmetry $K H = H K$ makes good sense in the interacting world -- it isn't just a gimmick limited to the special case of free fermions! We will present some examples of $K$ at work for interacting systems later on.

However, some practitioners might want to ignore interactions and deal only with Hamiltonians $H$ of the one-body form given in (\ref{eq:BdG}). With that proviso, one has the option of reducing the formalism of second quantization to the simplified setting offered by first quantization. Here for brevity, we confine ourselves to the case of number-conserving systems ($\Delta = 0$), referring to the end of Section \ref{sect:phsym-gapless} for the case of systems with pair condensation ($\Delta \not= 0$). The reduction step then amounts to just replacing the second-quantized Hamiltonian $H$ by the first-quantized Hamiltonian $h$.

In the reduced framework of first quantization, there is no room for the operation of exchanging creation and annihilation operators ($a \leftrightarrow a^\dagger$). All that remains visible of the particle-hole symmetry $K$ is the operation $\Gamma$. Here we wish to highlight that, in order for $K$ to be a symmetry of the second-quantized Hamiltonian, $\Gamma$ has to invert the sign of the first-quantized Hamiltonian $h$ of the non-interacting system:
\begin{equation}\label{eq:pseudo-Gamma}
    \Gamma h = - h \Gamma .
\end{equation}
In fact, for the cosine band of (\ref{eq:H-cosband}) we had $\Gamma t(k) \Gamma = t(k+\pi) = - t(k)$ (and an analogous sign reversal happens also for the Bogoliubov-deGennes Hamiltonian (\ref{eq:hBdG}) of the Kitaev chain). More generally, by comparing (\ref{eq:2QMH}) and (\ref{eq:H-phK}) we see that the symmetry condition $K H K^{-1} = H$ amounts to
\begin{equation*}
    \Gamma_{V_+ \to V_-} A = - D \Gamma_{V_+ \to V_-} \,, \quad
    \Gamma_{V_+ \to V_-} B = - C \Gamma_{V_- \to V_+} \,, \quad {\rm etc.},
\end{equation*}
which is none other than (\ref{eq:pseudo-Gamma}) in block-decomposed form.

If our notation is too abstract and unfamiliar, we can offer help by presenting the same calculation in standard physics notation: what leads from Eq.\ (\ref{eq:pseudo-Gamma}) to $K H = + H K$ is that the antilinear exchange $a \leftrightarrow a^\dagger$ produces a second sign change,
\begin{equation}\label{eq:oneline}
    \sum a_k^\dagger h_{kk^\prime}^{\vphantom{dagger}} \, a_{k^\prime}^{\vphantom{\dagger}} \rightarrow \sum a_k^{\vphantom{\dagger}} \overline{h_{kk^\prime}^{\vphantom{\dagger}}}\, a_{k^\prime}^\dagger = - \sum a_{k^\prime}^\dagger h_{k^\prime k}^{\vphantom{\dagger}} \, a_k^{\vphantom{\dagger}} \qquad (\mathrm{Tr} \, h = 0) ,
\end{equation}
to cancel the first sign change from $\Gamma h = - h \Gamma$. (Of course, we again had to assume that $\overline{h_{kk^\prime}} = h_{k^\prime k}$ are the matrix elements of a self-adjoint operator.)

A standard scenario for $\Gamma$ to arise exists in bipartite systems with sublattices $A$ and $B$. In fact, if the first-quantized free-fermion Hamiltonian $h$ is odd with respect to the sublattice decomposition $V = V_A \oplus V_B$ (i.e.\ if $h$ exchanges the summands $V_A$ and $V_B$), then it anti-commutes with the linear operation $\Gamma :\; \psi_A + \psi_B \mapsto \psi_A - \psi_B$ which reverses the sign of the single-particle wave function on the sites of the $B$-sublattice while keeping it the same on the $A$-sublattice.

In summary, from the one-line computation (\ref{eq:oneline}) we see that $\Gamma h = - h \Gamma$ makes for a particle-hole symmetry $K H = H K$ (with $K = c \circ \Gamma$) of the many-particle system at half filling. This then is the justification for the prominence of so-called chiral ``symmetries'' in the research area of topological quantum matter: in the special case of free fermions, the linear pseudo-symmetry $\Gamma h = - h \Gamma$ of the first-quantized Hamiltonian entails an antilinear true symmetry $K H = + H K$ of particle-hole type for the second-quantized Hamiltonian. Ending this subsection on a historical note, we mention that one of the main motivations for \cite{HHZ} to envisage a second anti-unitary symmetry generator (beyond that of time reversal) was to incorporate into the Tenfold Way certain random-matrix models \cite{Ver94} of chiral fermions in non-Abelian gauge backgrounds.

\subsection{Example 3: charge conjugation of the Dirac field}
\label{sect:C-Dirac}

To complete this expository account, we take our third example from relativistic quantum field theory: for Dirac fermions one has a (variant of) particle-hole symmetry which is known as charge-conjugation symmetry, $C$. Using the present language, we can describe it as follows.

In first quantization, a relativistic Dirac particle with mass $m$ and momentum $p$ obeys the dynamics of $\mathrm{i} \partial_t \psi = h \psi$ with Hamiltonian
\begin{equation}\label{eq:Dirac-h}
     h = \beta m + \sum \alpha_l \, p_l \,, \quad p_l = \frac{1} {\mathrm{i}} \, \frac{\partial}{\partial x_l} \qquad (\hbar = c = 1)
\end{equation}
in the absence of electromagnetic fields. The Dirac matrices are commonly chosen to be
\begin{equation*}
    \beta = \Bigg( \begin{array}{cc} \mathbf{1} &0 \\
    0 & - \mathbf{1} \end{array} \Bigg), \quad
    \alpha_l = \Bigg( \begin{array}{cc} 0 &\sigma_l \\
    \sigma_l &0 \end{array} \Bigg) \quad (l = 1, 2, 3) .
\end{equation*}
Given this choice, the charge-conjugation operator $\Gamma$ in the position-space representation is the involution ($\Gamma^2 = 1$)
\begin{equation*}
    \Gamma :\; V \to V , \quad \psi \mapsto \mathrm{i}\beta \alpha_2 \overline{\psi} \,, \quad V = \mathbb{C}^4 \otimes L^2(\mathbb{R}^3) .
\end{equation*}
Here (as almost always in the present paper), $\overline{\psi}$ means the complex conjugate of $\psi$. Thus charge conjugation $\Gamma$ in the first-quantized theory is antilinear.

The Dirac Hamiltonian $h$ in Eq.\ (\ref{eq:Dirac-h}) features eigenstates with positive energy above $m$ and negative-energy eigenstates below $-m$; these span $V_+$ and $V_-$ respectively. It is easy to see that $\Gamma$ anti-commutes with $h$ and therefore exchanges the eigenstates of positive and negative energy. Thus $\Gamma$ is a map $\Gamma : \; V_\pm \to V_\mp\,$.

By what was explained in the previous sections, the presence of the \emph{antilinear} operation $\Gamma$ in first quantization gives rise to
a \emph{linear} operation $c \circ \Gamma \equiv C$ on the Fock space $\mathcal{F} = \bigwedge(V_+ \oplus V_-^\ast)$ of the second-quantized theory. The latter commutes with the second-quantized Hamiltonian $H$ (for Dirac matter in zero electromagnetic field) and is \emph{unitary}.  Known as charge-conjugation symmetry, $C$ stabilizes the Dirac sea of filled negative-energy states and maps single-electron states to single-positron states.

In the presence of a fixed electromagnetic background, $C$ ceases to be a symmetry of the Dirac matter alone; it continues to be a symmetry of the full theory including the electromagnetic field, which  transforms as $C :\; (E,B) \mapsto (-E,-B)$.

For later use (in Section \ref{sect:LLL}), we append how charge conjugation is realized for Dirac fermions carrying electric charge $q$ in $2 + 1$ space-time dimensions. In that lowered space dimension, an irreducible spinor has but two components, and the Hamiltonian (including an electromagnetic field with gauge potential $A$) can be presented as
\begin{equation*}
     h = \sigma_3 m - q A_0 + \sum_{l=1}^2 \sigma_l \left( \frac{1} {\mathrm{i}} \, \frac{\partial}{\partial x_l} - q A_l \right) .
\end{equation*}
With this choice of first-quantized Hamiltonian, charge conjugation acts on the Hilbert space $V = \mathbb{C}^2 \otimes L^2(\mathbb{R}^2)$ as the antilinear operator
\begin{equation}\label{eq:C-2plus1}
    \Gamma :\; \psi \mapsto \sigma_1 \overline{\psi} .
\end{equation}
We observe that $- \Gamma h(A) = h(-A) \Gamma$. Of course, charge conjugation $C = c \circ \Gamma$ in the second-quantized theory is again linear (like in $3+1$ dimensions).

Also for later use, let us record that time reversal is realized as
\begin{equation}\label{eq:T-2plus1}
    T :\; \psi \mapsto \mathrm{i}\sigma_2 \overline{\psi} .
\end{equation}
Notice that time-reversal invariance for Dirac fermions in $2 + 1$ dimensions is not compatible with the presence of a mass term. This circumstance is related to the so-called parity anomaly \cite{Redlich,NiemiSemenoff}.

\subsection{Particle-hole symmetries in path-integral language}
\label{sect:path-int}

Up to now, we have been looking at particle-hole symmetries in the operator formalism of quantum mechanics, and for most of the paper we shall stick to that formalism. Yet, a complementary and fruitful perspective can be gained by taking a short excursion to see how particle-hole symmetries are realized in the path integral.

With a view toward our underlying theme of unitary and anti-unitary symmetries, we do not address the Euclidean (or imaginary-time) path integral but rather its variant in real time as the proper tool to compute the unitary quantum dynamics. For simplicity, we still look at the special case of a charge-conserving system with first-quantized Hamiltonian $h$ given in (\ref{eq:1QMH}) and second-quantized one-body Hamiltonian given in (\ref{eq:2QMH}). (The generalization to superconductors and interacting systems is not a problem.)

The real-time path integral for the time-evolution trace $\mathrm{Tr}\, \mathrm{e}^{- \mathrm{i} \tau H/\hbar}$ is then governed by the action functional (here let $\hbar = 1$, and put the summation convention in force)
\begin{equation}\label{eq:PI-action}
    S = \int_{0}^{\tau} |dt| \left( \mathrm{i} \xi_k \partial_t \eta^k - \xi_k h_{\;\;k^\prime}^k \eta^{k^\prime} \right) ,
\end{equation}
where $\xi_k(\tau) = - \xi_k(0)$ respectively $\eta^k(\tau) = - \eta^k(0)$ are fermion modes that quantize as creation resp.\ annihilation operators for $k > 0$ (corresponding to single-particle states in $V_+$) and as annihilation resp.\ creation operators for $k < 0$ (states in $V_-$). As usual, the path-integral action looks simpler than the second-quantized Hamiltonian (\ref{eq:2QMH}) expressed in terms of particles and holes, but this simplicity is balanced by the subtleties of handling the (proper analog of the) Feynman propagator.

We now ask how a particle-hole symmetry $K$ can be detected by inspection of the action functional (\ref{eq:PI-action}). As a warm-up, let us review how time-reversal symmetry $T$ is verified from the real-time path integral. Assuming that $T^2 = +1$, we adopt a basis of $T$-real fermion modes:
\begin{equation*}
    T \xi_k(t) = \xi_k(\tau - t) , \quad T \eta^k(t) = \eta^k(\tau-t) .
\end{equation*}
When applied to the path-integral action, the time-reversal operation sends $\xi_k$ to $T \xi_k$ and $\eta^k$ to $T \eta^k$, while taking the complex conjugate of the coefficients:
\begin{equation*}
    S \stackrel{T}{\longrightarrow} \int_{0}^{\tau} |dt| \, \xi_k(\tau - t) \Big( \overline{\delta_{\;\;k^\prime}^k \mathrm{i} \partial_t - h_{\;\; k^\prime}^k} \Big) \eta^{k^\prime}(\tau-t) .
\end{equation*}
Introducing $t^\prime = \tau - t$, and using $\overline{\mathrm{i} \partial_t} = \mathrm{i} \partial_{t^\prime}$ and $|dt| = |dt^\prime|$, we see that $S$ is $T$-invariant if and only if the matrix elements of $h$ are real, which is exactly the condition in a $T$-real basis for $h$ to be time-reversal symmetric. (Of course, the path integral is $\int \mathrm{e}^{\mathrm{i} S / \hbar}$, so quantum transition amplitudes and the time-evolution trace do change under $T$.)

We are now going to carry out the same calculation for the case of a particle-hole symmetry $K$. To begin, we fix a good basis of fermion modes so that
\begin{equation}\label{eq:K-on-xi}
    K \xi_k(t) = \eta^{k^\prime}(\tau-t) \Gamma_{k^\prime k} \,, \quad
    K \eta^k(t) = \Gamma^{k k^\prime} \xi_{k^\prime}(\tau-t) ,
\end{equation}
where $\Gamma_{k^\prime k} \equiv \overline{\Gamma_{\;\;k}^{k^\prime}}$ are the matrix elements of the antilinear operator $K = c \circ \Gamma : \; V \to V^\ast$, and $\Gamma^{k k^\prime} \equiv \overline{\Gamma_{\; \;k^\prime}^k}$ are the matrix elements of the inverse. As we shall see in a moment, the antilinear transformation (\ref{eq:K-on-xi}) leaves the action $S$ invariant. At first encounter, however, one may wonder why our operator $K$ of a particle-hole transformation inverts the arrow of time ($t \to \tau - t$), as one might regard that characteristic property as a prerogative of time reversal. To see why, notice that the particle-hole symmetry (as any good symmetry) must already be visible from the equation of motion for the quantum dynamics. Obtained by varying the action (\ref{eq:PI-action}) with respect to $\xi_k\,$,
\begin{equation}\label{eq:EOM4eta}
    (\delta_{\;\;k^\prime}^k \mathrm{i} \partial_t - h_{\;\; k^\prime}^k) \eta^{k^\prime} = 0 ,
\end{equation}
that equation features the first-quantized Hamiltonian $h$. So, if our particle-hole trans\-formation $K$ is a symmetry, any solution $t \mapsto \eta(t)$ of Eq.\ (\ref{eq:EOM4eta}) must transform to another solution, $t \mapsto K \eta(t)$. Now we know that $K = c \circ \Gamma$ acts in first quantization as $\Gamma$, which anti-commutes with $h$. Therefore the required effect (of solutions particle-hole transforming to solutions) takes place if and only if $\Gamma$ \emph{anti}-commutes with the operator $\mathrm{i} \partial_t - h$ of the dynamics. In order for that to happen, we need $\partial_t \to -\partial_t$ since $\Gamma$ is a linear operator ($\Gamma \mathrm{i} = + \mathrm{i} \Gamma$) and $\Gamma h = - h \Gamma$. We are thus led to the transformation law
\begin{equation}
    (\Gamma \eta)^k (t) = \Gamma_{\;\;k^\prime}^k \eta^{k^\prime}(\tau - t) .
\end{equation}
Given this, the second formula in (\ref{eq:K-on-xi}) follows from $K = c \circ \Gamma$ by applying the Fr\'echet-Riesz isomorphism, $c :\; \Gamma_{\;\;k^\prime}^k \eta^{k^\prime} \mapsto \overline{\Gamma_{\; \;k^\prime}^k} c(\eta^{k^\prime})$ and $c(\eta^k) = \xi_k\,$. The first formula then is a consequence of $K^2 = {\bf 1}$.

\smallskip\noindent\textit{Lemma}. The antilinear particle-hole transformation $K = c \circ \Gamma$ given by Eq.\ (\ref{eq:K-on-xi}) is a symmetry of the action functional $S$ in Eq.\ (\ref{eq:PI-action}) if and only if the linear transformation $\Gamma$ anti-commutes with the first-quantized Hamiltonian $h$.

\smallskip\noindent\textit{Proof}. We apply the formulas (\ref{eq:K-on-xi}) to the path-integral action:
\begin{equation*}
    S \to {}^K \! S = \int_{0}^{\tau} |dt| \, \eta^{l}(\tau - t) \Gamma_{l k} \Big( \overline{\delta_{\;\;k^\prime}^k \mathrm{i} \partial_t - h_{\;\; k^\prime}^k} \Big) \Gamma^{k^\prime l^\prime} \xi_{l^\prime}(\tau-t) .
\end{equation*}
Now from $\Gamma h \Gamma = - h$ and $h^\dagger = h$ we have
\begin{equation*}
    \Gamma_{lk} \overline{h_{\;\;k^\prime}^k} \Gamma^{k^\prime l^\prime} = \overline{\Gamma_{\;\;k}^l h_{\;\;k^\prime}^k \Gamma_{\;\; l^\prime}^{k^\prime}} = - \overline{h_{\;\;l^\prime}^l}
    = - h_{\;\;l}^{l^\prime} ,
\end{equation*}
and by again introducing $\tau-t$ as the new integration variable, we obtain
\begin{equation*}
    {}^K \! S = \int_{0}^{\tau} |dt| \, \eta^l(t) \left( \delta_{\;\;l}^{l^\prime} \mathrm{i} \partial_t + h_{\;\; l}^{l^\prime} \right) \xi_{l^\prime}(t) .
\end{equation*}
Finally, we switch the order of the Grassmann variables $\eta^l(t)$ and
$\xi_{l^\prime}(t)$, and we do partial integration for the summand containing $\partial_t\,$. Since both of these operations produce a sign change, we retrieve the original action: ${}^K \! S = S$. Thus $S$ is $K$-invariant, as claimed.

\smallskip\noindent\textit{Remark 1}. The substitution $t \mapsto \tau - t$  in (\ref{eq:K-on-xi}) indicates that our particle-hole symmetry $K = c \circ \Gamma$ is a relative of time-reversal symmetry $T$. Let it be emphasized, however, that $K$ and $T$ are markedly different in first quantization: $K$ (as $\Gamma$) is linear and anti-commutes with the first-quantized Hamiltonian $h$, whereas $T$ is antilinear and commutes with $h$. (Of course, both are antilinear and commute with the Hamiltonian $H$ in second quantization.) A revealing instance of this kinship will be presented in Section \ref{sect:LLL}, where we will learn that the two types of time inversion ($T$ versus $K$) get exchanged under fermionic particle-vortex duality in $2+1$ dimensions.

\smallskip\noindent\textit{Remark 2}. The attentive reader will have noticed that our particle-hole transformation $K$ bears some algebraic similarity with the operation $CT$ on relativistic quantum fields. However, it would be rather superficial and naive to make the identification $K \stackrel{?}{=} CT$. Indeed, the condensed-matter object $K$ knows nothing about special relativity (much less about charge conjugation $C$ for Dirac spinor fields), and in the examples above (Sects.\ \ref{sect:SSH}, \ref{sect:Kitaev}) $K$ shifts the (crystal) momentum $k$ by $\pi$, whereas $CT$ sends $k$ to $-k$.

\smallskip\noindent\textit{Remark 3}. Further to the comparison between $T$ and $K$, we note that $T$, when applied to quasi-particle excitations of a $T$-invariant ground state, keeps the electric charge of an excitation the same but reverses the electric current. In contradistinction, the situation for $K$ is the opposite (charge is reversed, current remains the same). Altogether, this means that $T$ as a symmetry tolerates the presence of an external electric field, whereas $K$ tolerates an external magnetic field. Thus $T$ and $K$ are complementary, and both are needed to encompass the full phenomenology encountered in condensed-matter systems.

\section{Gapless systems}\label{sect:gapless}

Let us now turn to the more challenging case of gapless systems. To give an example, in field theory one knows that domain walls between different vacua give rise to chiral fermions and zero modes. Similarly, the condensed-matter community has learned that topological insulators host protected zero modes or gapless states at their boundaries. Motivated by these and other examples of interest, we here take a look at particle-hole symmetries for systems with zero modes.

For that purpose, the decomposition of the single-particle Hilbert space $V = V_+ \oplus V_-$ by the conduction and valence subspaces is augmented to
\begin{equation*}
    V = V_+ \oplus V_0 \oplus V_- \,,
\end{equation*}
where $V_0$ denotes the subspace of zero (or close-to-zero) modes of the single-particle Hamiltonian; without too much loss, we assume the dimension of $V_0$ to be finite. In that expanded setting, the Fock space of the many-particle system becomes
\begin{equation*}
    \mathcal{F} = {\textstyle{\bigwedge}}(V_+ \oplus V_0 \oplus V_-^\ast) \cong {\textstyle{\bigwedge}}(V_0) \otimes {\textstyle{\bigwedge}} (V_+ \oplus V_-^\ast) .
\end{equation*}
A particle-hole transformation will be a map $K :\; \mathcal{F} \to \mathcal{F}$ that preserves the two factors of $\mathcal{F}$. Earlier, in Section \ref{sect:gapped}, we described $K$ as an antilinear transformation of the second factor $\bigwedge(V_+ \oplus V_-^\ast)$. Our task in the present section is to give a description of the other part, $K : \; \bigwedge(V_0) \to \bigwedge(V_0)$; the difference is that now there might be little gain from working with a Fock space built on a free-fermion ground state that occupies half of $V_0$ and leaves the other half empty. Indeed, there exist systems such as the half-filled lowest Landau level (see Section \ref{sect:LLL}) which are very far from the free-fermion limit and have strongly correlated ground states. With such cases in mind, we are going to set up particle-hole transformations on $\bigwedge(V_0)$ in another way.

To get started, we recall our basis-free notation
\begin{equation*}
    \varepsilon(v) :\; {\textstyle{\bigwedge}}^n(V_0) \to {\textstyle{\bigwedge}}^{n+1}(V_0) \quad {\rm and} \quad \iota(\varphi) :\; {\textstyle{\bigwedge}}^n(V_0) \to {\textstyle{\bigwedge}}^{n-1}(V_0)
\end{equation*}
for the basic Fock operators of particle creation ($v \in V_0$) and particle annihilation ($\varphi \in V_0^\ast$). The integer $n \geq 0$ here is the number of particles occupying zero modes (or states of very low energy with respect to the Fermi level). We also recycle the notation $c :\; V_0 \to V_0^\ast$, $v \mapsto \langle v \,\vert\, \cdot \rangle_{V_0}$ for the Fr{\'e}chet-Riesz isomorphism, with $\langle \cdot \vert \cdot \rangle_{V_0}$ resulting from the Hermitian scalar product of $V$ by the orthogonal projection $V \to V_0\,$.

For the sake of clear language, we will presently introduce the notion of  particle-hole \emph{conjugation}, which is distinct from a particle-hole \emph{transformation} (to be introduced later). Our choice of terminology is to indicate that the former is a canonical operation, while the latter needs input by a specific Hamiltonian. We begin by declaring how particle-hole conjugation acts on the operators (not yet the states).

\smallskip\noindent\emph{Definition.} Denoted by $A \mapsto A^\flat$, \emph{particle-hole conjugation} of operators is defined by
\begin{equation}\label{eq:PHT}
    \varepsilon(v)^\flat = \iota \big( c(v) \big) , \quad
    \iota(\varphi)^\flat = \varepsilon \big( c^{-1}(\varphi) \big),
\end{equation}
when acting on the basic Fock operators; it extends to the Clifford algebra of many-body operators as the antilinear algebra automorphism determined by Eqs.\ (\ref{eq:PHT}).

\medskip\noindent\emph{Remark.} Particle-hole conjugation coincides with the operation ``dagger'' of taking the Hermitian conjugate:
\begin{equation*}
    \varepsilon(v)^\dagger = \iota \big( c(v) \big) , \quad
    \iota(\varphi)^\dagger = \varepsilon \big( c^{-1}(\varphi) \big) ,
\end{equation*}
when restricted to the basic Fock operators. It differs from Hermitian conjugation in that it preserves the operator order: $(AB)^\flat = A^\flat B^\flat$, whereas $\dagger$ reverses it: $(AB)^\dagger = B^\dagger A^\dagger$.

\subsection{Fr\'echet-Riesz on $\bigwedge(V_0)$}

Introducing the operation of particle-hole conjugation by its action on the Fock operators is an unsatisfying short cut to a more fundamental approach. Indeed, particle-hole conjugation should be defined, in the first instance, as a transformation of the Fock space, and that Fock-space transformation should induce the transformation (\ref{eq:PHT}) by conjugation. Therefore we now address the question as to how the relations (\ref{eq:PHT}) lift to Fock space. In keeping with our general style, we construct the lift in a basis-free manner, by using nothing but the invariant structures at hand.

Particle-hole conjugation lifted to the Fock space $\bigwedge(V_0)$ (for $V_0$ of finite dimension) is best viewed as a composition of two maps. To describe the first of these, we observe that the given Hermitian scalar product on $V_0$ determines a Hermitian scalar product on each exterior power $\bigwedge^n(V_0)$. Thus the Fr{\'e}chet-Riesz correspondence $c :\; V_0 \to V_0^\ast$ has a canonical extension to Fock space as
\begin{equation}\label{eq:FRn}
    c_n :\; {\textstyle{\bigwedge}}^n(V_0) \to {\textstyle{\bigwedge}}^n (V_0)^\ast , \quad \Psi \mapsto \langle \Psi \vert \,\cdot \rangle_{\wedge(V_0)} \,,
\end{equation}
for all $n \geq 0$. We reiterate that $c_n$ is antilinear. For a particle-number changing operator $L :\; \bigwedge^m(V_0) \to \bigwedge^n(V_0)$, it determines the Hermitian conjugate $L^\dagger :\; \bigwedge^n(V_0) \to \bigwedge^m(V_0)$ by $L^\dagger = c_m^{-1} \circ L^\ast \circ c_n\,$. [Recall that any linear transformation $L :\; A \to B$ of vector spaces has a canonical adjoint $L^\ast :\; B^\ast \to A^\ast$ on the dual vector spaces by $(L^\ast \beta)(a) = \beta(L a)$.]

\subsection{Wedge isomorphism}

To define the other factor featuring in the Fock-space lift of particle-hole conjugation, we make use of the assumption $\mathrm{dim}\, V_0 \equiv N < \infty$. Fixing some unitary element $\Omega \in \bigwedge^N (V_0)$ of top degree (physically speaking: a normalized Slater determinant filling all single-particle states in $V_0$) we define a $\mathbb{C}$-linear bijection
\begin{equation}\label{eq:wedge-iso}
    \omega_n :\; {\textstyle{\bigwedge}}^n(V_0)^\ast \to {\textstyle{\bigwedge}}^{N-n}(V_0) , \quad \Phi \mapsto \omega_n \Phi ,
\end{equation}
by requiring for all test vectors $\Psi^\prime \in \bigwedge^n (V_0)$ the equality
\begin{equation}\label{eq:4.3}
    (\omega_n \Phi) \wedge \Psi^\prime = \Phi(\Psi^\prime) \, \Omega_n \,,
\end{equation}
where $\Omega_n \in \bigwedge^N(V_0)$ is defined recursively by
\begin{equation}\label{eq:4.4}
    \Omega_0 \equiv \Omega \,, \quad
    \Omega_n = (-1)^{N-n}\, \Omega_{n-1} \quad (n = 1, 2, \ldots, N) .
\end{equation}
We refer to $\omega_n$ as the wedge isomorphism. It induces another map on operators, say $L \mapsto L^\wedge$. For $L :\; \bigwedge^m(V_0) \to \bigwedge^n (V_0)$ the wedge-transformed operator $L^\wedge :\; \bigwedge^{N-n} (V_0) \to \bigwedge^{N-m}(V_0)$ is given by $L^\wedge = \omega_m \circ L^\ast \circ \omega_n^{-1}\,$. Because the map $L \mapsto L^\wedge$ invokes the canonical adjoint $L^\ast$, it reverses the order of operators just like Hermitian conjugation does. Note also that our induced map on operators does not depend on the choice of phase for $\Omega$. Indeed, that phase factor drops out because it occurs twice in $\omega_m \circ L^\ast \circ \omega_n^{-1}$, once as itself and once as its inverse.

The varying sign factors in (\ref{eq:4.4}) have been inserted in order for the following statement to take its simplest (i.e., sign-free) form.

\smallskip\noindent\textit{Lemma}. For all basic Fock operators $L = \varepsilon(v)$ and $L = \iota(\varphi)$ one has $L^\wedge = L$.

\smallskip\noindent\textit{Proof}. Let $L = \varepsilon(v) :\; \bigwedge^\bullet(V) \to \bigwedge^{\bullet\, + 1}(V)$ be a particle-creation operator. In this case the claimed equality $L^\wedge = L$ is equivalent to
\begin{equation*}
    \varepsilon(v) \circ \omega_n = \omega_{n-1} \circ \varepsilon(v)^\ast \quad (n = 1, \ldots, N) .
\end{equation*}
To verify that relation, we apply both sides to $\Phi \in \bigwedge^n (V_0)^\ast$ and test the result against a vector $\Psi \in \bigwedge^{n-1}(V_0)$. Starting on the left-hand side, we then move to the right-hand side by the following computation:
\begin{eqnarray*}
    &&\varepsilon(v) \, \omega_n \Phi \wedge \Psi =
    (-1)^{N-n} \omega_n \Phi \wedge \varepsilon(v) \Psi = \cr
    &=& (-1)^{N-n} \Phi( \varepsilon(v) \Psi ) \, \Omega_n
    = (\varepsilon(v)^\ast \Phi)(\Psi) \, \Omega_{n-1} =
    \omega_{n-1} \, \varepsilon(v)^\ast \Phi \wedge \Psi .
\end{eqnarray*}
Since $\Phi$ and $\Psi$ are arbitrary, this already proves the desired relation.

The case of a particle-annihilation operator $L = \iota(\varphi) :\; \bigwedge^\bullet(V_0) \to \bigwedge^{\bullet\, - 1}(V_0)$ is not much different: for $\Phi \in \bigwedge^{n-1}(V_0)^\ast$ and $\Psi \in \bigwedge^n (V_0)$ one shows that the relation
\begin{equation*}
    \iota(\varphi)\,\omega_{n-1} \Phi \wedge \Psi = \omega_n \iota(\varphi)^\ast \Phi \wedge \Psi \quad (n = 1, \ldots, N)
\end{equation*}
holds by using that $\iota(\varphi) \, \omega_{n-1} \Phi \wedge \Psi = (-1)^{N-n} \omega_{n-1} \Phi \wedge \iota(\varphi) \Psi$.

\subsection{Lifting particle-hole conjugation to Fock space}
\label{sect:lifting-PH}

With the wedge isomorphism $\omega_n$ and the Fr\'echet-Riesz isomorphism $c_n$ in hand, we concatenate them to form a map
\begin{equation}\label{eq:Xi-n}
    \Xi_n :\; {\textstyle{\bigwedge}}^n(V_0) \stackrel{c_n}{\longrightarrow} {\textstyle{\bigwedge}}^n (V_0)^\ast \stackrel{\omega_n}{\longrightarrow} {\textstyle{\bigwedge}}^{N-n} (V_0) .
\end{equation}
Note that $\Xi_n = \omega_n \circ c_n$ is antilinear by the presence of the antilinear factor $c_n\,$, since the wedge isomorphism $\omega_n$ is linear. We denote the entire family of maps $\Xi_n$ by
\begin{equation}
    \Xi :\; {\textstyle{\bigwedge}}(V_0) \to {\textstyle{\bigwedge}}(V_0) .
\end{equation}
Note further that the induced map $L \mapsto \Xi L\, \Xi^{-1}$ preserves the operator order; indeed, each of the maps induced by the $c_n$ and $\omega_n$ (namely, $L \mapsto L^\dagger$ and $L \mapsto L^\wedge$) reverses it.

\smallskip\noindent\textit{Lemma}. The operator $\Xi$ is a lift of particle-hole conjugation $L \mapsto L^\flat$ to Fock space, i.e.,
\begin{equation}\label{eq:flat-lifted}
    L^\flat = \Xi \circ L \circ \Xi^{-1} .
\end{equation}

\noindent\textit{Proof}. Both $L \mapsto L^\flat$ and $L \mapsto \Xi L \, \Xi^{-1}$ are antilinear algebra automorphisms. Therefore, they agree if they agree on the set of Clifford algebra generators $\varepsilon(v)$ and $\iota(\varphi)$.

Hence consider $L \equiv \varepsilon(v) :\; \bigwedge^{n-1}(V_0) \to \bigwedge^n(V_0)$. Since Hermitian conjugation is an involution, one has $\varepsilon(v) = (\varepsilon(v)^\dagger)^\dagger = c_n^{-1} (\varepsilon(v)^\dagger)^\ast c_{n-1}\,$. It follows that
\begin{equation*}
    \Xi_n \, \varepsilon(v) \, \Xi_{n-1}^{-1} = \omega_n (c_n \varepsilon(v) c_{n-1}^{-1}) \, \omega_{n-1}^{-1} = \omega_n (\varepsilon(v)^\dagger)^\ast \omega_{n-1}^{-1} = (\varepsilon(v)^\dagger)^\wedge .
\end{equation*}
Now $\varepsilon(v)^\dagger = \iota(c(v))$ and from the previous Lemma we know that $\iota(\varphi)^\wedge = \iota(\varphi)$. Thus $\Xi \, \varepsilon(v)\, \Xi^{-1} = \iota(c(v))^\wedge = \varepsilon(v)^\dagger = \varepsilon(v)^\flat$, as desired. In the same manner one shows that $\Xi \, \iota(\varphi) \, \Xi^{-1} = \iota(\varphi)^\dagger = \iota(\varphi )^\flat$. This completes the proof. $\square$

\smallskip\noindent\textit{Example 1.}
Let us compute $\Xi :\; \bigwedge^\bullet (V_0) \to \bigwedge^\bullet (V_0)$ for the simple case of the Hermitian vector space $V_0 = \mathbb{C}^2$ with orthonormal basis $\{ e_+ , e_- \}$, dual basis $\{ f_+ , f_- \}$ of $V_0^\ast$, and $\Omega = e_+ \wedge e_-\,$. Taking $\Phi = f_+$ and $\Psi^\prime = v \in V_0$ in Eq.\ (\ref{eq:4.3}) we get
\begin{equation*}
    \omega_1(f_+) \wedge v = f_+ (v)\, \Omega_1 = - \Omega \, f_+(v) = e_- \wedge e_+ f_+(v) = e_- \wedge v ,
\end{equation*}
where $\Omega_1 = - \Omega$ due to Eq.\ (\ref{eq:4.4}) for $N - n = 2 - 1$. It follows that $\omega_1(f_+) = e_-$ and hence
\begin{equation*}
    \Xi \, e_+ = (\omega_1 \circ c_1)(e_+) = \omega_1 (f_+) = e_- \,.
\end{equation*}
The other cases can be done in the same way. Altogether, we find
\begin{equation*}
    \Xi \cdot 1 = e_+ \wedge e_- \,, \quad
    \Xi \, e_+ = e_- \,, \quad \Xi \, e_- = - e_+ \,,
    \quad \Xi (e_+ \wedge e_-) = - 1 .
\end{equation*}
Note that $\Xi^2 = - \mathbf{1}_{\wedge(V_0)}$ in the present example.
For future reference, note also that although $\Xi$ differs from the anti-unitary operator $T$ of time reversal in general, $\Xi$ agrees with $T$ on restriction to $\bigwedge^1(V_0) = V_0$ for $V_0 = \mathbb{C}^2_{\rm spin}$.

\smallskip\noindent\textit{Example 2.}
Next, we consider the case of a Hermitian vector space $V_0$ of any finite dimension and claim that
\begin{equation}\label{eq:4.10-new}
    \Xi^2 = (-1)^{N(N-1)/2}\, \mathbf{1}_{\wedge(V_0)} \quad (N = \mathrm{dim}\, V_0) .
\end{equation}
This is seen as follows. First of all, observe that $\Xi^2$ commutes with all basic Fock operators $\varepsilon(v)$ and $\iota(\varphi)$. Indeed,
\begin{equation*}
    \Xi^2 \, \varepsilon(v) \, \Xi^{-2} = \Xi \, \iota(c(v)) \, \Xi^{-1} = \varepsilon(c^{-1} \circ c(v)) = \varepsilon(v) ,
\end{equation*}
and likewise for $\iota(\varphi)$. Since the Clifford algebra generated by the operators $\varepsilon(v)$ and $\iota(\varphi)$ acts irreducibly on Fock space, it follows that $\Xi^2$ is a multiple of the unit operator.

To compute the constant of proportionality, we apply $\Xi^2$ to the Fock vacuum, $1 \in \bigwedge^0(V_0)$. From the definition (\ref{eq:4.3}) of the wedge isomorphism, we immediately see that $\Xi \cdot 1 = \Omega$. Now
\begin{equation*}
    (\Xi^2 \cdot 1) \wedge \Omega = (\Xi \, \Omega) \wedge \Omega = \langle \Omega \vert \Omega \rangle_{\wedge^N(V_0)} \, \Omega_N = \Omega_N \,,
\end{equation*}
since $\Omega$ has norm one. By carrying out the recursion of Eq.\ (\ref{eq:4.4}), we get
\begin{equation*}
    \Omega_N = \Omega_{N-1} = - \Omega_{N-2} = - \Omega_{N-3} = ... = (-1)^{N(N-1)/2}\, \Omega_0\,.
    \end{equation*}
Hence $(\Xi^2 \cdot 1) \wedge \Omega = \Omega_N = (-1)^{N(N-1)/2}\, \Omega$ and by dropping $\Omega$ we arrive at Eq.\ (\ref{eq:4.10-new}).

\smallskip\noindent\textit{Remark}.
By suppressing the varying sign factors from (\ref{eq:4.4}) and inserting $\Phi = c_n \Psi$ into Eq.\ (\ref{eq:4.3}), we could have given a slightly different definition (cf.\ \cite{OHRMT}) of $\Xi$ as
\begin{equation}\label{eq:Hodge}
    (\Xi \Psi) \wedge \Psi^\prime \stackrel{?}{=} \langle \Psi \vert \Psi^\prime \rangle_{\wedge(V_0)}\, \Omega \,,
\end{equation}
which would have made it look the same as the definition of the Hodge star operator on differential forms in Riemannian geometry but for the feature that $\Xi$ is antilinear. [On the down side, the omitted sign factors would result in $\Xi \, \varepsilon(v) \, \Xi^{-1} \equiv \varepsilon(v)^\flat = \pm \varepsilon(v)^\dagger$.] $\square$

\begin{figure}
    \begin{center}
        \epsfig{file=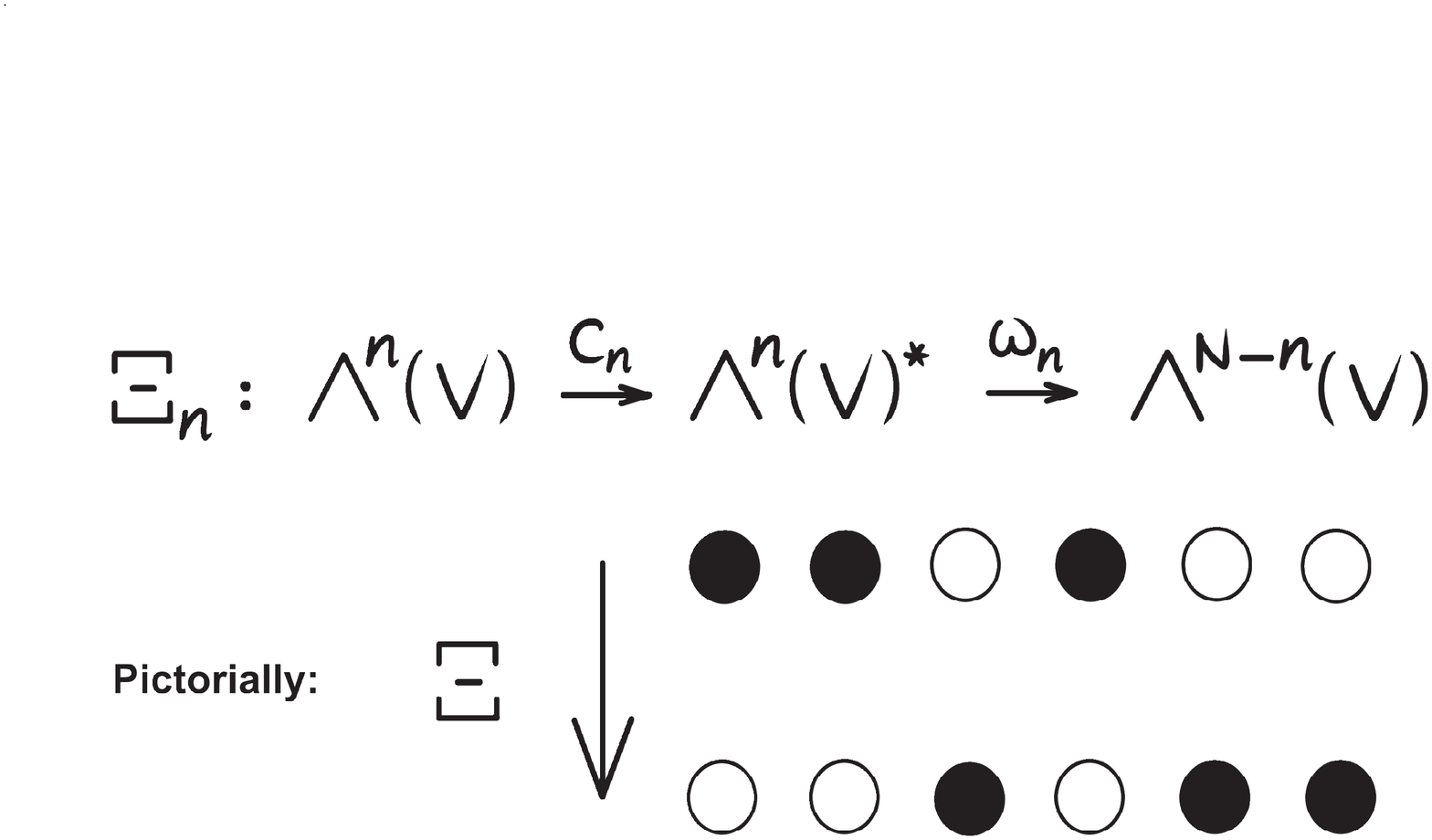,height=8cm}
    \caption{Transforming $n$ particles into $n$ holes, the antilinear operator $\Xi_n$ of particle-hole conjugation is invariantly defined as the concatenation of the Fr\'echet-Riesz isomorphism $c_n$ with the wedge isomorphism $\omega_n\,$. In the occupation-number representation with respect to any orthonormal single-particle basis, it takes filled orbitals to empty ones and vice versa. Consequently, particle-hole conjugation can never be a symmetry of any Fermi liquid.} \label{fig:phlifted}
    \end{center}
\end{figure}

Having constructed particle-hole conjugation $\Xi :\; \bigwedge^n(V_0) \to \bigwedge^{N-n}(V_0)$ in a basis-free manner, we hasten to add that the action of $\Xi$ is very simple in the occupation-number representation of the Fock space $\bigwedge(V_0)$ for any orthonormal basis $\{\vert k\rangle\}$ of $V_0\,$. In fact, if $n_k \in \{ 0 , 1 \}$ are the occupation numbers or eigenvalues of $\varepsilon(\vert k \rangle) \iota(\langle k \vert) = a_k^\dagger a_k^{\vphantom{\dagger}}\,$, then one infers from the definitions (\ref{eq:FRn}, \ref{eq:wedge-iso}, \ref{eq:Xi-n}) that $\Xi$ just has the effect (up to an overall sign) of exchanging $n_k \leftrightarrow 1-n_k\,$, i.e., empty orbitals are particle-hole conjugated to filled orbitals and vice versa (cf.\ Fig.\ \ref{fig:phlifted}).

\subsection{Example: lowest Landau level (Girvin, 1984)}
\label{sect:Girvin}

The present notion of particle-hole conjugation made a pioneering appearance in \cite{Girvin} for the two-dimensional electron gas of the quantum Hall effect, as follows. Assuming the limit of a very strong magnetic field, one may project the Hilbert space of a single electron to the lowest Landau level, say $V_0\,$. In the symmetric gauge for a magnetic field of strength $B = - |B|\, dx \wedge dy$ (cf.\ \cite{vanEnk} for the minus sign), the latter is realized by complex-analytic functions $\psi$ of $z = (x + \mathrm{i}y) / \ell$ where $x, y$ are Cartesian coordinates for the Euclidean plane (more generally: local coordinates on a Riemann surface) and $\ell$ is the magnetic length, $\ell = \sqrt{\hbar / |e B|}$. The Hermitian scalar product on $V_0$ is given by
\begin{equation*}
    \langle \psi \vert \psi^\prime \rangle_{V_0} = \int d\mu(z)\, \overline{\psi(z)}\, \psi^\prime (z) , \quad d\mu(z) = \frac{| dx\, dy |}{2\pi\ell^2}\, \mathrm{e}^{-\vert z \vert^2 / 2} .
\end{equation*}
For a disk-shaped finite-size system threaded by a total magnetic flux of $N$ flux quanta, $V_0$ is spanned by the functions $z^j$ for $j = 0, 1, \ldots, N-1$ and thus has dimension $N$.

The many-electron wave function for the state $\Omega \in \bigwedge^N (V_0)$ of total filling can be expressed as a Vandermonde determinant,
\begin{equation*}
    \Omega(z_1, z_2, \ldots, z_N) = \mathcal{N}^{-1/2} \prod_{1 \leq i < j \leq N} (z_i - z_j) ,
\end{equation*}
with a suitable normalization constant $\mathcal{N}$. Using $\Omega$, the operation of particle-hole con\-jugation $\bigwedge^n(V_0) \to \bigwedge^{N-n} (V_0)$, $\Psi \mapsto \Xi_n \Psi$, of Eq.\ (\ref{eq:Hodge}) takes the concrete form
\begin{eqnarray}
    &&(\Xi_n \Psi)(z_{n+1}, \ldots, z_N) \nonumber \\
    &&= \int \prod_{j=1}^n d\mu(z_j) \, \overline{\Psi(z_1, \ldots, z_n)} \, \Omega(z_1, \ldots, z_n, z_{n+1}, \ldots, z_N) . \label{eq:4.6}
\end{eqnarray}
Thus a complex-analytic function of $n$ variables is particle-hole conjugated to a complex-analytic function of $N-n$ variables. (Actually, $\Psi$ is a totally skew polynomial of degree $\leq N-1$ in each variable, and so is $\Xi_n \Psi$.) Eq.\ (\ref{eq:4.6}) amounts to the same as (\ref{eq:Hodge}). To see that, one takes the exterior product of (\ref{eq:4.6}) with an arbitrary test function $\Psi^\prime \in \bigwedge^n(V_0)$ and observes that $\Psi^\prime \otimes c_n \Psi$ applied to $\Omega \in \bigwedge^N(V_0)$ gives $\Omega$ times $\langle \Psi \vert \Psi^\prime \rangle_{\wedge(V_0)}\,$.

Let us stress again the obvious fact that $\Xi_n \, \lambda = \overline{\lambda} \, \Xi_n$ for $\lambda \in \mathbb{C}$. This characteristic feature of particle-hole conjugation on Fock space is inevitable so long as the Hermitian scalar product is the only invariant structure available for its construction.

\subsection{Particle-hole symmetries for gapless free fermions}
\label{sect:phsym-gapless}

We recall that the operator $\Xi$ of particle-hole conjugation, when expressed in the occupation-number representation for any orthonormal basis $\vert k \rangle$ of single-particle states, simply reverses the occupancy ($n_k \to 1-n_k$). As a consequence, $\Xi$ can \emph{never} be a true symmetry of any physical system near the free-fermion limit! Indeed, $\Xi$ takes a Fermi-liquid state occupying single-particle states of momentum $| {\bf k} | < k_F$ to the Fermi-liquid state occupying the states of momentum $| {\bf k} | > k_F$. This implies that $\Xi$ cannot leave the many-body ground state invariant unless that state is highly correlated.

Thus, to realize a particle-hole symmetry for free fermions, particle-hole conjugation $\Xi$ alone does not suffice. From our treatment of gapped systems in Section \ref{sect:gapped}, we may anticipate what other structure is needed: an involution $\Gamma$ that anti-commutes with the first-quantized Hamiltonian $h$. To formulate the optimal statement covering super\-conductors as well as particle-number conserving systems, we begin with a preparation.

In the lemma stated below, we employ the terminology of \emph{Weyl-ordered} one-body operator. By this we mean an operator that lies in the span of the skew-symmetrized quadratic operators
\begin{equation*}
    \varepsilon(v) \iota(\varphi) - \iota(\varphi) \varepsilon(v) , \quad \varepsilon(v) \varepsilon(v^\prime), \quad \iota(\varphi) \iota(\varphi^\prime) ,
\end{equation*}
where $v , v^\prime$ range through $V_0$ and $\varphi , \varphi^\prime$ range through $V_0^\ast$.

\smallskip\noindent\textit{Lemma.}
On Weyl-ordered one-body operators $A$ the particle-hole conjugate coincides with the skew-Hermitian adjoint:
\begin{equation}\label{eq:4.10}
    \Xi A \, \Xi^{-1} \equiv A^\flat = - A^\dagger .
\end{equation}

\smallskip\noindent\textit{Proof.}
Recall that $(AB)^\flat = A^\flat B^\flat$ preserves the operator order whereas $(AB)^\dagger = B^\dagger A^\dagger$ reverses it. Now for the elementary Fock operators $\alpha = \varepsilon(v)$ or $\alpha = \iota(\varphi)$ we also recall the equality $\alpha^\flat = \alpha^\dagger$. Using it, one computes for any Weyl-ordered one-body operator $A = \alpha \beta - \beta \alpha$ that
\begin{equation*}
    A^\flat = (\alpha \beta - \beta \alpha)^\flat = \alpha^\flat \beta^\flat - \beta^\flat \alpha^\flat = - (\beta^\dagger \alpha^\dagger - \alpha^\dagger \beta^\dagger) = - A^\dagger .
\end{equation*}
The statement then follows because both $\flat$ and $\dagger$ are antilinear. $\square$

From Eq.\ (\ref{eq:4.10}) we see that a second-quantized Hamiltonian $H$ which is Weyl-ordered, one-body, and self-adjoint, is odd with respect to particle-hole conjugation $\Xi$. Now any one-body Hamiltonian $H$ can be arranged to be Weyl-ordered by subtracting a constant, the trace of $H$ on $\bigwedge^1(V_0)$. Thus without loss of generality we may assume our self-adjoint one-body Hamiltonians $H$ to be particle-hole odd: $H^\flat = - H^\dagger = - H$. Clearly then, to arrive at a particle-hole symmetry $K H = + H K$, we need to compose $\Xi$ with a second transformation to compensate for the sign change due to $\Xi$.

Following Sect.\ \ref{sect:3.2.1} let $\Gamma :\; V \to V$ be an involution ($\Gamma^2 = \mathbf{1}_V$) that anti-commutes with the first-quantized free-fermion Hamiltonian $h$ acting on the single-particle Hilbert space $V$:
\begin{equation*}
    \Gamma h = - h \Gamma ,
\end{equation*}
where for the moment we assume particle number to be conserved. We further assume that $\Gamma$ is unitary or anti-unitary. In either case, the property of anti-commutation with $h$ implies that $\Gamma$ maps eigenstates of energy $E$ to eigenstates of energy $-E$:
\begin{equation*}
     h v = E v \quad \Longleftrightarrow \quad h \Gamma v = -E \Gamma v .
\end{equation*}
Thus if $V = V_+ \oplus V_0 \oplus V_-$ is the decomposition into the subspaces of positive, zero, and negative energy for $h$, then $\Gamma V_\pm = V_\mp$ and $\Gamma V_0 = V_0\,$. In particular, $\Gamma$ induces a family of maps on the Fock space of the zero modes:
\begin{equation}\label{eq:Gamma-n}
    \Gamma_n :\; {\textstyle{\bigwedge}}^n(V_0) \to {\textstyle{\bigwedge}}^n(V_0) , \quad v_1 \wedge \cdots \wedge v_n \mapsto \Gamma v_1 \wedge \cdots \wedge \Gamma v_n \,.
\end{equation}
Finally, we compose $\Gamma_n$ with $\Xi_n$ to arrive at a mapping
\begin{equation}\label{eq:C-n}
    K_n \equiv \Xi_n \circ \Gamma_n :\; {\textstyle{\bigwedge}}^n(V_0) \to {\textstyle{\bigwedge}}^{N-n}(V_0) .
\end{equation}
Writing $K :\; \bigwedge(V_0) \to \bigwedge(V_0)$ for short, we refer to such an operation $K$ as a particle-hole transformation [more precisely, as a ph-transformation restricted to the Fock space of zero modes -- the full transformation $K :\; \mathcal{F} \to \mathcal{F}$ for $\mathcal{F} = \bigwedge(V_0) \otimes \bigwedge(V_+ \oplus V_-^\ast)$ includes $K :\; \bigwedge(V_+ \oplus V_-^\ast) \to \bigwedge(V_+ \oplus V_-^\ast)$ as defined in (\ref{eq:C-Fock})].

Assuming the anti-commutation relation $\Gamma h = - h \Gamma$, it now follows quickly from $\Xi H = - H \Xi$ that the particle-hole transformation $K = \Xi \circ \Gamma$ is a symmetry of the second-quantized Hamiltonian: $K H = H K$. We stress again that $K$ is antilinear if $\Gamma$ is linear (as will usually be the case in examples from condensed matter physics).

To finish this subsection, let us comment briefly on the case of superconductors in the free-fermion approximation. In that situation, one does not have a first-quantized Hamiltonian $h$ in the standard sense of particle-number conserving systems; as a substitute, one considers the so-called Bogoliubov-deGennes Hamiltonian \cite{NNB,deGennes}, which is an operator on the space $V \oplus V^\ast$ of fields (or Nambu spinors \cite{Nambu}):
\begin{equation}\label{eq:hBdG}
    h_{\rm BdG} = \Bigg( \begin{array}{cc} h &\Delta \\
    \Delta^\dagger & - h^\ast \end{array} \Bigg),
\end{equation}
where $h^\ast \in \mathrm{End}(V^\ast)$ is the canonical adjoint of $h \in \mathrm{End}(V)$, and $\Delta \in \mathrm{Hom}(V^\ast,V)$ is a skew-symmetric map induced from pairing interactions by the formation of a condensate $\langle \psi \psi \rangle$ of Cooper pairs. The second-quantized Hamiltonian $H$ in (\ref{eq:BdG}) then has the particle-hole symmetry $K H = H K$ if $\Gamma h = - h \Gamma$ and $\Gamma \Delta = - \Delta \Gamma^\ast$. An example of such a Hamiltonian $H$ is that of the Kitaev chain (\ref{eq:kitaev}). The present discussion based on $V = V_+ \oplus V_0 \oplus V_-$ becomes relevant for the Kitaev chain with boundary, where symmetry-protected zero modes appear at the end(s) of the chain.

\section{Interacting systems}\label{sect:interact}

So far, we have illustrated the notion of particle-hole symmetry at the example of free-fermion Hamiltonians such as the Su-Schrieffer-Heeger model and the Kitaev chain. Yet, having introduced particle-hole conjugation $\Xi$ and particle-hole transformations $K$ on Fock space, we are now ready to turn to some examples of particle-hole symmetry taken from the realm of interacting many-fermion systems. These will include the Hubbard model at half filling, quantum spin chains in the Haldane phase, and last but not least, the half-filled lowest Landau level.

\subsection{Example 1: Hubbard model}

For notational simplicity, we here take the space dimension to be $d = 1$; the generalization to an arbitrary dimension will be obvious.

Consider an $\mathrm{SU}(2)$-spin doublet of fermions with spin quantum number $s = \pm 1/2$, hopping among the sites $x \in \mathbb{Z}$ of a chain. The Hamiltonian of the Hubbard model is made of the kinetic energy of hopping plus an on-site repulsive interaction ($U > 0$):
\begin{equation}\label{eq:Hubbard}
    H = - \sum_{x, \,s} (t \, a_{x+1,\,s}^\dagger a_{x,\,s}^{\vphantom{\dagger}} + \bar{t} \, a_{x,\,s}^\dagger a_{x+1,\,s}^{\vphantom{\dagger}}) + U \sum_x Q_x^2 \,.
\end{equation}
Assuming the chemical potential to be that for half filling, $Q_x$ is the Weyl-ordered charge (or excess charge relative to half filling) at the site $x :$
\begin{equation*}
    Q_x = \sum_s a_{x,\,s}^\dagger a_{x,\,s}^{\vphantom{\dagger}} - 1
    = \frac{1}{2} \sum_s \left[ a_{x,\,s}^\dagger \,, \, a_{x,\,s}^{\vphantom{\dagger}} \right] .
\end{equation*}
The kinetic part of (\ref{eq:Hubbard}) consists of two copies ($s = \pm 1/2$) of the cosine band (\ref{eq:H-cosband}). We already know that it has the particle-hole symmetry (\ref{eq:ph-Kitaev}), acting diagonally on the spin quantum number. The same particle-hole transformation is also a symmetry of the two-body interaction term. This is because
\begin{equation*}
     K Q_x K^{-1} = \frac{1}{2} \sum_s \left[ (-1)^x a_{x,\,s}^{ \vphantom{\dagger}} \, , (-1)^x a_{x,\,s}^\dagger \right] = - Q_x \,.
\end{equation*}
Thus the excess charge $Q_x$ is particle-hole odd, and its square $Q_x^2$ is particle-hole even.

Now let $\Psi_0$ be a ground state of $H$ at half filling. It follows from $K H = H K$ that the particle-hole transform $K \Psi_0$ is another ground state of $H$ at half filling. If that ground state is unique, then $K \Psi_0 \in \mathbb{C} \cdot \Psi_0\,$, meaning that $\Psi_0$ is particle-hole symmetric. States with excitation energy $E$ and excess charge $q$ (relative to the ground state) are mapped by $K$ to excited states with the same energy $E$ and the opposite charge $-q$.

Note that while the Hubbard Hamiltonian (\ref{eq:Hubbard}) satisfies $K H = H K$ irrespective of the filling fraction, the particle-hole transformation $K$ maps a ground state below half filling to a ground state above half filling and vice versa. In other words, the particle-hole symmetry of the ground state is broken when the chemical potential moves away from half filling. Note also that $K$ remains a symmetry of the Hamiltonian even when the parameters $t$ and $U$ are made space-dependent or random; also, nonlocal charge-charge interactions are compatible with the particle-hole symmetry $K$.

\subsection{Example 2: Heisenberg spin chain}
\label{sect:Heisenberg}

At half filling and for a large Hubbard coupling $U > 0$, all low-energy states of the Hamiltonian (\ref{eq:Hubbard}) have particle occupation number $n_x = 1$ at every site. (Indeed, any vacant or doubly occupied site costs a large energy $U$.) In that limit, the low-energy physics of the Hubbard model is that of a quantum spin chain with Hilbert space $\bigotimes_x \mathbb{C}_x^2$ acted upon by spin operators
\begin{equation*}
    S_x^i = \frac{1}{2} \sum_{s s^\prime} a_{x,\,s}^\dagger (\sigma^i)_{s s^\prime} \, a_{x,\,s^\prime}^{\vphantom{\dagger}} \quad (i = 1,2,3).
\end{equation*}
The effective Hamiltonian for the spin chain is well known to be
\begin{equation}\label{eq:Bethe}
    H_S = J \sum_x \bigg( \sum_i S_x^i S_{x+1}^i - 1/4 \bigg) ,
\end{equation}
with antiferromagnetic coupling $J = 2 |t|^2 / U > 0$.

The quantum spin chain (\ref{eq:Bethe}) is known as the (Bethe-ansatz solvable) Heisenberg spin chain. By its origin from the half-filled Hubbard model with particle-hole symmetry $K$, it shares the same ph-symmetry $K$, a property that can be verified directly from
\begin{equation*}
    K S_x^i K^{-1} = \frac{1}{2} \sum_{s s^\prime} (-1)^x a_{x,\,s}^{\vphantom{\dagger}} \overline{(\sigma^i)}_{s s^\prime} (-1)^x a_{x,\,s^\prime}^\dagger = - S_x^i \,.
\end{equation*}
Here we should make the remark that the Heisenberg chain is usually regarded as a system with time-reversal symmetry:
\begin{equation*}
     T S_x^i T^{-1} = \frac{1}{2} \sum_{s s^\prime} (-1)^{1/2-s} a_{x,\,-s}^\dagger \overline{(\sigma^i)}_{s s^\prime} (-1)^{ 1/2-s^\prime} a_{x,\,-s^\prime}^{\vphantom{\dagger}} = - S_x^i \,.
\end{equation*}
This duplicity of symmetry denominations should not come as much of a surprise, as one can see from Eq.\ (\ref{eq:4.10-new}) for $N = 2$ that particle-hole conjugation $\Xi$ and hence $K = \Xi \circ \Gamma$ acts like time reversal $T$ in what is known as the ``Mott limit'', i.e.\ after restriction to the space of states of the spin chain.

\subsection{Symmetry-protected zero modes for $G = \mathrm{U}(1) \times \mathbb{Z}_2^K$} \label{sect:SPT-AIII}

For a first application of particle-hole symmetry in gapped interacting systems, let us consider a set of $n$ gapped Su-Schrieffer-Heeger (SSH) chains coupled by electron-electron interactions. We require the Hamiltonian of the interacting system to have the symmetry group $G = \mathrm{U}(1)_Q \times \mathbb{Z}_2^K$, i.e., charge $Q$ is conserved and there is a particle-hole symmetry $K$ (such systems are said to be of symmetry type $A\mathrm{I\!I\!I}$). For example, if the coupled system consists of $n = 2$ chains corresponding to the two spin components of spinful electrons, the Hamiltonian might be that of the one-dimensional Hubbard model with staggered hopping.

Our interest now is in semi-infinite chains at half filling with particle-hole symmetry $K = \Xi \circ \Gamma$. We label the atomic sites by $x = 0, 1, 2, \ldots$ and the chains by an index $\nu = 1, 2, \ldots, n$. If the gapped bulk system carries a non-trivial topological invariant, then by the principle of bulk-boundary correspondence one expects a number of localized zero modes to appear at the edge $x = 0$, making the zero-energy ground state degenerate.

First, we review the situation for free fermions. Let
\begin{equation}\label{eq:0-mode}
    \alpha = \sum_{x \geq 0} \sum_{\nu = 1}^n z_{x,\nu}\, a_{x,\nu} \quad (z_{x,\nu} \in \mathbb{C})
\end{equation}
be a zero mode, i.e.\ a solution of $[H , \alpha] = 0$. Any such zero mode of the second-quantized Hamiltonian $H$ stems from the space $V_0$ of zero modes of the first-quantized Hamiltonian $h$. Since $h$ anticommutes with the involution $\Gamma$, the latter acts on $V_0$ to decompose it into two eigenspaces $V_0^\mp = \mathrm{ker}(\Gamma \pm \mathbf{1})$. For any zero mode drawn from $V_0^+$ or $V_0^-$ the corresponding operator $\alpha$ in Eq.\ (\ref{eq:0-mode}) is particle-hole even ($K \alpha = + \alpha^\dagger K$) or particle-hole odd ($K \alpha = - \alpha^\dagger K$), respectively.{}

Next, we augment the free-fermion Hamiltonian $H$ by perturbations coupling the different zero modes. Constrained by particle-hole symmetry and Hermiticity, all zero-mode couplings in the free-fermion approximation must be of the form
\begin{equation*}
    \alpha_+^\dagger \alpha_-^{\vphantom{\dagger}} + \alpha_-^\dagger \alpha_+^{\vphantom{\dagger}} = K ( - \alpha_+^{\vphantom{\dagger}} \alpha_-^\dagger - \alpha_-^{\vphantom{\dagger}} \alpha_+^\dagger ) K^{-1}
\end{equation*}
with particle-hole even $\alpha_+$ and particle-hole odd $\alpha_-$. The result of such a perturbation are two modes of equal positive energy (w.r.t.\ the perturbed many-body ground state) and opposite charge, namely $\beta_{\rm p} = \alpha_+ + \alpha_-$ and $\beta_{\rm h} = \alpha_+^\dagger - \alpha_-^\dagger = K \beta_{\rm p} K^{-1}$, since
 \begin{equation*}
    \alpha_+^\dagger \alpha_-^{\vphantom{\dagger}} + \alpha_-^\dagger \alpha_+^{\vphantom{\dagger}} = \left( [ \beta_{\rm p}^\dagger , \beta_{\rm p}^{\vphantom{\dagger}} ] + [ \beta_{\rm h}^\dagger , \beta_{\rm h}^{\vphantom{\dagger}}] \right) / 4 .
\end{equation*}
Thus zero modes can be gapped out pairwise, by always canceling one ph-even mode against one ph-odd mode. This mechanism of gapping out zero modes has the difference $\mathrm{dim}\, V_0^+ - \mathrm{dim}\, V_0^- \in \mathbb{Z}$ as a free-fermion topological invariant. Taking the principle of bulk-boundary correspondence for granted, we thus obtain the  $\mathbb{Z}$-classification of $A\mathrm{I\!I\!I}$-symmetry protected free-fermion topological phases in one space dimension \cite{Prodan-Schuba}.

Second, we turn on electron-electron interactions. The particle-hole symmetry $K H = H K$ of the Hamiltonian still implies that $K$ acts on the space $\mathrm{ker} H \equiv \mathcal{V}_0$ of ground states: if $\Psi_0 \in \mathcal{V}_0$ then $K \Psi_0 \in \mathcal{V}_0\,$. Now for a generic system of interacting chains, there exists no compelling reason for the ground-state degeneracy to be any higher than that forced by the particle-hole symmetry; hence we expect $\mathrm{dim} \mathcal{V}_0 \leq 2$ generically. By that token, we obtain four different possibilities, and hence a fourfold classification, depending on just the sign of $K^2 \vert_{ \mathcal{V}_0} = \pm 1$ and on whether $K$ is even or odd with respect to fermion parity, $K (-1)^F = \pm (-1)^F K$:
\begin{enumerate}
\item The ground state is unique; i.e., $\Psi_0$ and $K\Psi_0$ are linearly dependent, which requires $K^2 \vert_{ \mathcal{V}_0} = + 1$ and $K (-1)^F = (-1)^F K$.
\item If $K^2 \vert_{ \mathcal{V}_0} = - 1$ and $K (-1)^F = (-1)^F K$, one has a pair of linearly independent ground states (a so-called Kramers pair) of the same fermion parity.
\item $K^2 \vert_{ \mathcal{V}_0} = +1$ and $K (-1)^F = - (-1)^F K$. Here the two ground states $\Psi_0$ and $K \Psi_0$ have opposite fermion parity; thus $K$ acts as a kind of supercharge.
\item Same as (iii) but with $K^2 \vert_{\mathcal{V}_0} = -1$.
\end{enumerate}

\noindent{\it Remark 1.} Consider the case of $n$ identical SSH chains, each contributing one localized zero mode for the boundary at $x = 0$. Then in the limit of decoupled chains we have $\mathrm{dim} \mathcal{V}_0 = \mathrm{dim} \bigwedge\!(\mathbb{C}^n) = 2^n$. Let us further assume that $\Xi$ and $\Gamma$ commute, so that $K^2 = (\Xi \circ \Gamma)^2 = \Xi^2$ and $K^2 \vert_{\mathcal{V}_0} = (-1)^{n(n-1)/2}$ from Eq.\ (\ref{eq:4.10-new}). Under these assumptions, the cases (i)--(iv) above will occur for $n = 0 , 2, 1, 3$ (mod $4\mathbb{Z}$), respectively, when the ground-state degeneracy $2^n$ is lifted by turning on interactions between the chains. Mathematically speaking, $K$ determines on the complex Grassmann algebra $\bigwedge(\mathbb{C}^n)$ an
\begin{itemize}
\item even real structure for $n = 0$ (mod $4\mathbb{Z}$),
\item even quaternionic structure for $n = 2$ (mod $4\mathbb{Z}$),
\item odd real structure for $n = 1$ (mod $4\mathbb{Z}$),
\item odd quaternionic structure for $n = 3$ (mod $4\mathbb{Z}$),
\end{itemize}
where even and odd refers to the decomposition $\bigwedge(\mathbb{C}^n) = \bigwedge^{\rm even} \oplus \bigwedge^{\rm odd}$ by fermion parity. Such structures are invariant under deformation by interactions.

\smallskip\noindent{\it Remark 2.} The classification above is known \cite{MFM2015,QKS2016} as the $\mathbb{Z}/4\mathbb{Z}$ classification of $A\mathrm{I\!I\!I}$-protected fermionic topological phases in one space dimension. As we shall see, our treatment provides a new perspective on the celebrated Haldane phase as a realization of case (ii) above -- a perhaps surprising link, which we now address for our first major application of particle-hole symmetry in interacting systems.

\section{From free fermions to the Haldane phase}\label{sect:FF2Haldane}

Quantum spin chains have a long history going back to the early days of quantum mechanics. The translation-invariant $\mathrm{SU}(2)$-symmetric chain with spin $S = 1/2$ and antiferromagnetic coupling was shown by Lieb--Schultz--Mattis \cite{LSM1961} to possess a unique ground state with gapless excitations in the limit of an infinite chain. The gapless feature is known \cite{AffleckLieb} to persist for higher spins $S = 3/2$, $5/2$, $7/2$, etc. On the other hand, arguing on the basis of the non-linear sigma model approximation valid for large $S$, Haldane \cite{HaldaneConj} made the Nobel-Prize winning conjecture that antiferromagnetic quantum spin chains with \emph{integer} $S$ have a disordered ground state with a finite energy gap for excitations. Now over the years, it transpired that there is a twist to that story: for the gapped chains with integer spin $S$ it makes a difference whether $S$ is even or odd. In fact, chains with even $S$ are topologically trivial, whereas for odd $S$ there exists a kind of hidden topological order, which becomes manifest for chains with boundary by the appearance of zero-energy end modes with fractionalized spin. In the case of $S = 1$ these features are easily verified for the exactly solvable Affleck--Kennedy--Lieb--Tasaki (AKLT) Hamiltonian \cite{AKLT} and its matrix-product ground state.

\subsection{Haldane phase as an SPT phase}

It is understood nowadays that the AKLT spin chain represents a large class of topological gapped one-dimensional systems referred to collectively as the ``Haldane phase''. Unlike, say, fractional quantum Hall systems, the Haldane phase does not have a topological order that is stable with respect to just any deformation of the Hamiltonian; it is stable only under a class of deformations restricted by symmetry and is therefore called a symmetry-protected topological phase.

Now the question arises: protection by which symmetry? Thinking within the restricted class of systems with only spin degrees of freedom, Pollmann--Berg--Turner--Oshikawa \cite{PBTO} argued that the Haldane phase is protected by any one of three symmetries: (i) time reversal, (ii) the dihedral group $\mathbb{Z}_2 \times \mathbb{Z}_2$ or (iii) a spatial inversion symmetry. This proposal for symmetry protection was found to be unsatisfactory in view of earlier work by Anfuso--Rosch \cite{AR2007}, who had shown that, by introducing local charge fluctuations, one can deform ground states in the Haldane phase to trivial atomic product states while preserving time-reversal symmetry and the dihedral group. The Anfuso--Rosch objection was countered by Moudgalya--Pollmann \cite{MP2015} who demonstrated by numerical computation of the entanglement spectrum that the Haldane phase is protected by a bond-centered inversion symmetry even when local charge fluctuations are allowed. However, one expects the Haldane phase to be stable not only with respect to local charge fluctuations but also w.r.t.\ disorder that violates the inversion symmetry of Moudgalya--Pollmann. In order to fill that theory gap, this author argued some time ago \cite{MRZ-2016} that the Haldane phase is protected by $G = \mathrm{U}(1)_Q \times \mathbb{Z}_2^K$ with the particle-hole symmetry $K$ of Sect.\ \ref{sect:SPT-AIII}, which is tolerant of disorder. Soon after, an expanded version of that proposal appeared in a paper by Verresen--Moessner--Pollmann \cite{VMP2017}.

To be clear about what the statement is, we make a couple of definitions.

\smallskip\noindent\textit{Definition 1.} Two Hamiltonians of a bulk system without boundary are said to belong to the same (invertible) topological phase if they are connected by a homotopy, or continuous deformation, subject to the condition that every Hamiltonian along the path of the homotopy has a unique ground state with an energy gap for excitations.

\smallskip\noindent\textit{Definition 2.} A fermion Hamiltonian is said to be of symmetry class $A \mathrm{I\!I\!I}$ if it conserves the charge $Q$ and commutes with an operator $K$ of particle-hole transformation.

\smallskip\noindent\textit{Remark.} Here as always, by ``charge'' we mean the particle number or, equivalently, the total electric charge of a system of identical particles carrying charge. We mention in passing that Wang and Senthil (see, e.g., \cite{WS-AIII}) have considered a realization of our symmetry class $A\mathrm{I\!I\!I}$ by conserved spin projection $S_z$ (instead of $Q$) and time reversal $T$ (instead of $K$). In either case, the symmetry group is a direct product $G = \mathrm{U}(1) \times \mathbb{Z}_2\,$.

Now, by restricting the allowed homotopies to be within class $A\mathrm{I\!I\!I}$, we obtain the notion of $A\mathrm{I\!I\!I}$-symmetry protected topological phase. Our claim then is that the Haldane phase of antiferromagnetic quantum spin chains is connected by homotopy (in $A\mathrm{I\!I\!I}$) to a well-known and well-studied $A\mathrm{I\!I\!I }$-symmetry protected topological phase of free fermions (for the latter see, e.g., \cite{Prodan-Schuba}). The argument goes as follows.

\subsection{Two staggered Hubbard chains coupled by Hund's rule}

\begin{figure}
    \begin{center}
        \epsfig{file=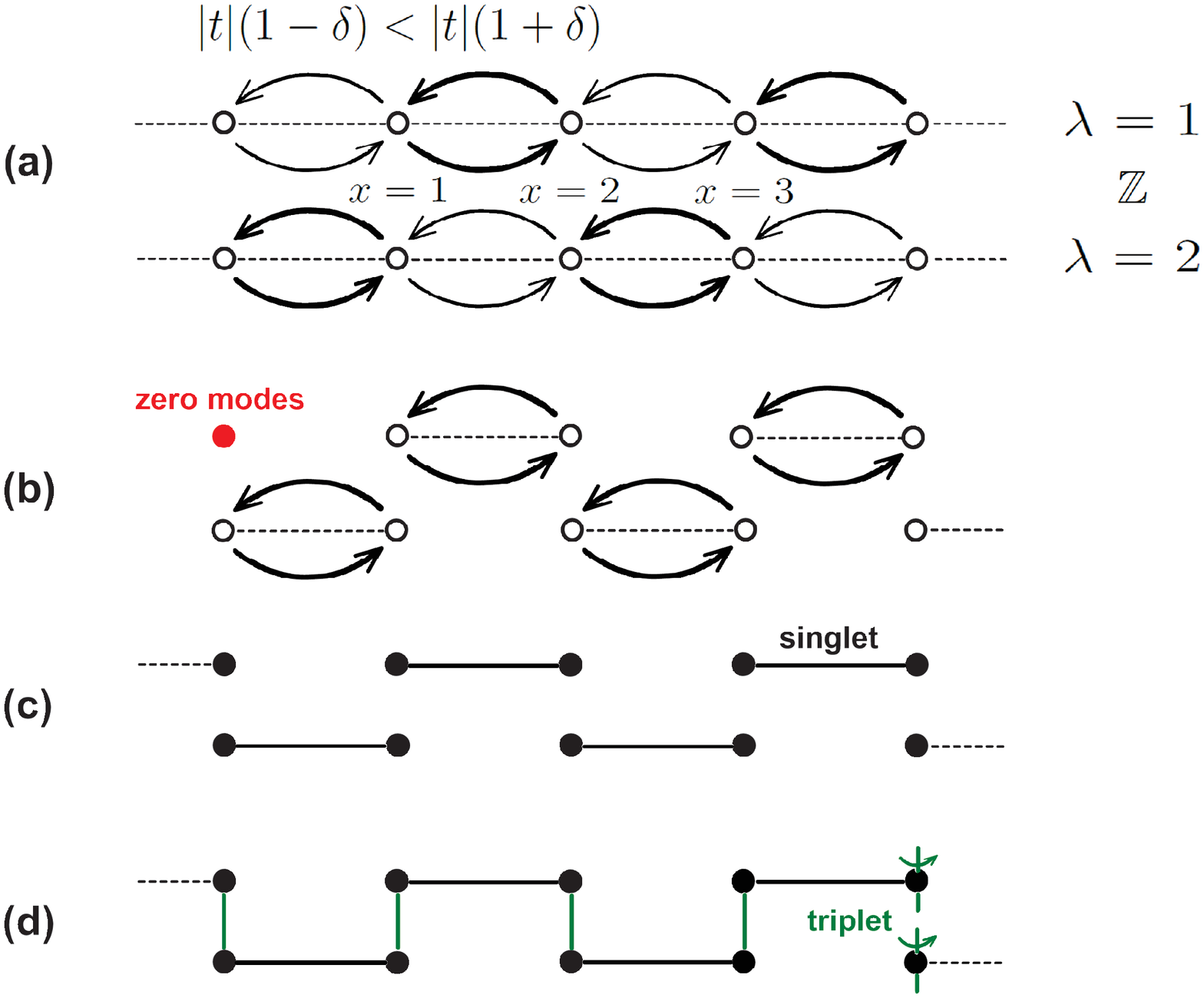,height=8cm}
    \caption{Adiabatic deformation taking two spinful SSH chains to the spin $S = 1$ antiferromagnetic Heisenberg chain. (a) Starting point: two uncoupled SSH chains ($\lambda = 1, 2$) for fermions with spin $S = 1/2$ (not shown), where the staggered hopping $|t|(1 \pm \delta)$ is shifted between the two chains. (b) First, the weak hopping $|t|(1-\delta)$ is tuned to zero. If the chain is truncated, zero modes (two for spin $1/2$) appear at the end. (c) Second, the intra-chain Hubbard coupling is turned on ($U \gg |t|$). At half filling, this results in exactly two charges per site. Spin singlets form between pairs of sites connected by nonzero hopping. (d) Third, the inter-chain Hund's rule coupling ($J \sim U$) is turned on, resulting in low-energy states with spin $S = 1$ on each site.}
    \label{fig:ssh2fdmh}
    \end{center}
\end{figure}

In a nutshell, our strategy is to set out from two Su-Schrieffer-Heeger (SSH) chains of spinful free fermions at half filling and subsequently introduce an intra-chain Hubbard interaction and an inter-chain Hund's rule coupling \cite{Affl-Hald}. By dialing up these interaction terms, we are able to deform (cf.\ Fig.\ \ref{fig:ssh2fdmh}) the free-fermion Hamiltonian to the Hamiltonian of the spin-$1$ antiferromagnetic Heisenberg chain, without leaving the symmetry class $A\mathrm{I\!I\!I}$ and without closing the energy gap.

As before, we label the atomic sites by $x \in \mathbb{Z}$ and the two chains by $\lambda = 1, 2$. The spin index is $s = \pm 1/2$. Our Hamiltonian consists of three terms:
\begin{eqnarray}
    H = &-& |t| \sum_{x,\lambda,s} \left( 1 + (-1)^{x+\lambda} \delta \right) \left( a_{x+1,\lambda s}^\dagger \, a_{x,\lambda s}^{\vphantom{\dagger}} + \mathrm{h.c.} \right) \nonumber \\ &+& U \sum_x \left( Q_{x,1}^2 + Q_{x,2}^2 \right) - V \sum_{x, i} S_{x,1}^i S_{x,2}^i \,. \label{eq:Ham-HH}
\end{eqnarray}
The hopping strength $|t| (1\pm \delta)$ alternates by an amount $2\delta |t| > 0$ along the chains and also between the chains. $U > 0$ is the parameter of a repulsive intra-chain Hubbard interaction expressed in terms of normal-ordered local charges
\begin{equation*}
    Q_{x,\lambda} = \frac{1}{2} \sum_s \left[ a_{x,\lambda s}^\dagger \,, a_{x,\lambda s}^{\vphantom{\dagger}} \right] .
\end{equation*}
The parameter $V > 0$ is a Hund's rule coupling between the two chains with local spin operators
\begin{equation*}
    S_{x,\lambda}^i = \frac{1}{2} \sum_{s s^\prime} a_{x,\lambda s}^\dagger \sigma_{s s^\prime}^i a_{x,\lambda s^\prime }^{\vphantom{\dagger}} \,.
\end{equation*}
We observe that the Hamiltonian (\ref{eq:Ham-HH}) has the particle-hole symmetry $K H = H K$ with $K a_{x,\lambda s}^\dagger = (-1)^x a_{x,\lambda s} K$. Assuming half filling (which amounts to two electrons per site, as there are four orbitals per site due to $\lambda = 1, 2$ and $s = \pm 1/2$), we are going to deform $H$ inside the four-dimensional parameter space of $|t|$, $\delta$, $U$, and $V$.

\subsection{Flat bands and valence-bond limit}

To begin the discussion, we turn off the interactions ($U = V = 0$). The system then consists of four decoupled SSH chains, counting spin. Our first goal is to determine the numerical topological invariant of that gapped free-fermion system. This is done most easily by taking the staggering parameter to $\delta = 1$, thereby sending the weak hopping $|t|(1-\delta)$ to zero. Now our fermions do nothing but hop between the two sites of some pair $(x,x^\prime)$ of adjacent sites; for fixed $\lambda \in \{1,2\}$ these are the pairs $(x,x^\prime) = (2m,2m + (-1)^\lambda)$, which are connected by a strong bond $|t|(1+\delta) = 2|t|$. If we then truncate the system to $x \geq 0$, so that $x = 0 $ becomes a boundary site, there are two zero modes $a_{0,\lambda s}$ ($s = \pm 1/2$) at the edge for the $\lambda = 1$ chain; cf.\ Figure \ref{fig:ssh2fdmh} (b). Thus the free-fermion topological invariant is $I = 2 \in \mathbb{Z}$ (or $I = -2$, depending on conventions) by the principle of bulk-boundary correspondence \cite{Prodan-Schuba,AMZ}.

The same conclusion is reached by inspecting the Bloch Hamiltonian $h(k)$ for the momentum-space representation of the period-2 translation-invariant bulk system. Taking the unit cell to encompass the pairs of sites that are connected by hopping on the chain $\lambda = 2$, we obtain the (first-quantized) flat-band Hamiltonians
\begin{equation*}
    h(k)_{\lambda=1} = - 2|t| \Bigg( \begin{array}{cc} 0 &\mathrm{e}^{\mathrm{i}k} \cr \mathrm{e}^{-\mathrm{i}k} &0 \end{array} \Bigg) , \quad h(k)_{\lambda=2} = - 2|t| \Bigg( \begin{array}{cc} 0 &1 \cr 1 &0 \end{array} \Bigg) .
\end{equation*}
Thus the spin-$1/2$ chain for $\lambda = 1$ has winding number one (from $\mathbb{R} / 2\pi\mathbb{Z} \ni k \mapsto \mathrm{e}^{\mathrm{i}k}$), while that for $\lambda = 2$ has winding number zero ($k \mapsto 1$). This means (see, for example, \cite{Prodan-Schuba}) that the system carries the topological invariant $\pm I = 1 + 1 = 2$.

Next, we crank up the intra-chain Hubbard interaction to a large value $U \gg |t|$. With that interaction in place, all low-energy states have exactly two fermions on every site $x$ (one for $\lambda = 1$ and $\lambda = 2$ each). From Section \ref{sect:Heisenberg} and Eq.\ (\ref{eq:Bethe}) the resulting low-energy physics is that of two uncoupled spin-$1/2$ chains with Hamiltonian
\begin{equation}\label{eq:H-spin}
    H_{\rm spin} = \frac{2|t|^2}{U} \sum_{x,\lambda}
    \left( 1 + (-1)^{x+\lambda} \delta \right)^2
    \bigg( \sum_i S_{x,\lambda}^i \, S_{x+1,\lambda}^i - 1/4 \bigg) .
\end{equation}
For simplicity, let us continue to assume the limit of maximal staggering, $\delta = 1$. Then the antiferromagnetic exchange coupling on the chain $\lambda$ is nonzero only between pairs $(x,x^\prime) \in (2\mathbb{Z}, 2\mathbb{Z} + (-1)^\lambda)$,  and the interacting ground state becomes a simple product of valence-bond singlets formed between those neighboring sites that are connected by nonzero antiferromagnetic exchange. The system now has charged excitations of high energy and spin excitations of low energy. To produce the latter, an energy $8 |t|^2 / U$ has to be spent to break up a valence bond.

\subsection{Deformation to the Heisenberg chain}

Up to now, we have been dealing with two uncoupled Hubbard chains. In our final step, we bring the Hund's rule coupling between the chains into play, raising it to a value $V \gg |t|^2 / U$. We expect this deformation to leave the energy gap still open \cite{footnote-DIS}. In fact, a large ferromagnetic coupling $V$ just pushes to high energy all the states containing spin singlets on the Hund's rule coupled sites. Its primary effect is to align the spins $1/2$ of the two fermions per site to form a composite spin $1$. Thus the two spin-$1/2$ chains get fused to a single spin-$1$ chain.

Our remaining task is to determine the low-energy effective Hamiltonian for the emerging spin-$1$ chain. To leading order in the small parameter $|t|^2 / (UV)$, the effective Hamiltonian is obtained by simply projecting the spin Hamiltonian (\ref{eq:H-spin}) to the spin-$1$ sector. We will perform this projection for a pair of neighboring sites $(x,x+1)$ with $x \in 2\mathbb{Z}$ (the case of $x \in 2\mathbb{Z}+1$ is identical), where antiferromagnetic exchange takes place only on the $\lambda = 2$ chain. We then need to calculate the expression
\begin{equation}\label{eq:project}
    (P_{x}^{S=1} \otimes P_{x+1}^{S=1}) \bigg( \sum\nolimits_i S_{x,\lambda=2}^i \otimes S_{x+1,\lambda=2}^i - 1/4 \bigg) (P_{x}^{S=1} \otimes P_{x+1}^{S=1}) ,
\end{equation}
with $P_x^{S=1}$ the projector on the spin-triplet states at $x$. Doing this calculation is an easy job of angular-momentum recoupling, as follows. Let
\begin{equation*}
    \psi_x = \Big( [1/2]_{x,\lambda=1} \otimes [1/2]_{x,\lambda=2} \Big)^{S=1} = P_x^{S=1} \psi_x
\end{equation*}
be the spin triplet made from the two spin-$1/2$ states $[1/2]_{x,\lambda}$ on the site $x$. The tensor product $\psi_x \otimes \psi_{x+1}$ decomposes into states $(\psi_x \otimes \psi_{x+1})^J$ of total angular momentum $J = 0, 1, 2$. On the other hand, let
\begin{equation*}
    \phi_{\lambda}^S = \Big( [1/2]_{x,\lambda} \otimes [1/2]_{x+1,\lambda} \Big)^S
\end{equation*}
be the spin singlet ($S = 0$) and triplet ($S = 1$) states formed by angular-momentum coupling of the two spin-$1/2$ states on the adjacent sites $(x,x+1)$ of the chain $\lambda$. Then a straightforward calculation using the pertinent Clebsch-Gordan coefficients yields
\begin{equation*}
    \begin{array}{l}
    (\psi_x \otimes \psi_{x+1})^{J=0} = - \frac{\sqrt{3}}{2}
    (\phi_{1}^{0} \otimes \phi_{2}^{0})^{J=0} - \frac{1}{2} (\phi_{1}^{1} \otimes \phi_{2}^{1})^{J=0} , \cr
    (\psi_x \otimes \psi_{x+1})^{J=1} = \frac{1}{\sqrt{2}}
    (\phi_{1}^{0} \otimes \phi_{2}^{1})^{J=1} + \frac{1}{\sqrt{2}} (\phi_{1}^{1} \otimes \phi_{2}^{0})^{J=1} , \cr
    (\psi_x \otimes \psi_{x+1})^{J=2} = (\phi_{1}^{1} \otimes \phi_{2}^{1})^{J=2} .
    \end{array}
\end{equation*}
Now the states $(\psi_x \otimes \psi_{x+1})^J$ diagonalize any $\mathrm{SU}(2)$-invariant interaction such as (\ref{eq:project}) between the spin-triplet states $\psi_x$ and $\psi_{x+1}\,$. Computing the
expectation values
\begin{equation*}
    h_J \equiv \left\langle (\psi_x \otimes \psi_{x+1})^J \Big\vert \sum S_{x,2}^i \, S_{x+1,2}^i - 1/4 \Big\vert (\psi_x \otimes \psi_{x+1})^J \right\rangle
\end{equation*}
in the basis of the angular-momentum re-coupled states $(\phi_1^\bullet \otimes \phi_2^\bullet)^J$, we obtain
\begin{equation*}
    h_0 = - 3/4, \quad h_1 = - 1/2, \quad h_2 = 0 .
\end{equation*}
After inclusion of the overall factor from (\ref{eq:H-spin}), these matrix elements are reproduced by the Hamiltonian
\begin{equation}\label{eq:H-eff1}
    H_{\rm eff} = \frac{2 |t|^2}{U} \sum_{x} \bigg( \sum_i \mathcal{S}_x^i \mathcal{S}_{x+1}^i - 1 \bigg) ,
\end{equation}
where $\mathcal{S}_x^i$ are the spin operators acting in the 3-dimensional space spanned by the spin-triplet states $\psi_x\,$. We observe that the projection to the spin-triplet sector has made the Hamiltonian translation-invariant with period 1 (shortening the initial period of 2).  The Hamiltonian (\ref{eq:H-eff1}) is the standard Heisenberg Hamiltonian for the quantum spin-$1$ chain with antiferromagnetic coupling. Haldane argued \cite{HaldaneConj} that it has a unique ground state with an energy gap for excitations, by appealing to the correspondence at long wavelengths with the $\mathrm{O}(3)$ nonlinear sigma model. Using current terminology, we say that the Hamiltonian (\ref{eq:H-eff1}) is in the Haldane phase.

The upshot is that we have deformed our gapped free-fermion Hamiltonian (\ref{eq:Ham-HH}) (with $U = V = 0$) to the gapped antiferromagnetic spin-$1$ Heisenberg Hamiltonian (\ref{eq:H-eff1}). Thus we have demonstrated an adiabatic connection between the Haldane phase and free fermions. In the wider perspective, we have identified the Haldane phase as part of the $A\mathrm{I\!I\!I}$-symmetry protected topological phase $I = \pm 2 \in \mathbb{Z} / 4\mathbb{Z}$ of interacting fermions in one space dimension (Sect.\ \ref{sect:SPT-AIII}).

\section{Half-filled lowest Landau level}\label{sect:LLL}

We now turn to a second major example illustrating the use of particle-hole symmetries in interacting condensed-matter systems: electrons in the lowest Landau level (LLL) at half filling (or filling fraction $\nu = 1/2$). That example differs from all our earlier ones in that the symmetry operation turns out to be plain (!) particle-hole conjugation $K = \Xi$ instead of a particle-hole transformation $K = \Xi \circ \Gamma$ (i.e., in the present instance $\Gamma$ is taken to be trivial, $\Gamma \equiv \mathbf{1}$). As we have emphasized, the plain ph-operator $\Xi$ cannot ever be a symmetry of any Fermi-liquid ground state (for the electrons). Nonetheless, Son \cite{DTSon} has recently proposed for $\nu = 1/2$ a particle-hole symmetric Fermi-liquid ground state built from so-called Dirac com\-posite fermions. It is an exciting story and lesson in many-body theory to learn how Son's proposal gets around the apparent obstruction.

\subsection{Symmetry under particle-hole conjugation}

The hallmark of quantum dynamics projected to the lowest Landau level (or any Landau level, for that matter) is that the kinetic energy of the charge carriers is totally quenched (if disorder or inhomogeneities in the background potential can be neglected), leaving no one-body component in the Hamiltonian of the bulk system. Now since the particle-hole conjugation operator $\Xi$ sign-inverts the local charge density ($\Xi Q_x \Xi^{-1} = - Q_x$) with respect to half filling, any residual two-body charge-charge (or current-current) interaction, in particular the Coulomb interaction, commutes with $\Xi$. What breaks the particle-hole conjugation symmetry $\Xi$ are three-body interactions (or energy-dependent two-body interactions induced by screening) and excitations into higher Landau levels. The latter are negligible in the limit of a very strong magnetic field $B$, where the cyclotron energy $\hbar \omega_c \propto |B|$ is much larger than the other energy scales of the problem.

In the sequel, we assume our Hamiltonian $H$ to be exactly particle-hole conjugation symmetric ($\Xi H = H \Xi$). Under that assumption, we might expect the ground state to be particle-hole conjugation symmetric ($\Xi \Psi_0 \in \mathbb{C} \Psi_0$) at half filling. If so, we face an immediate complication from the free-fermion perspective: since $\Xi$ exchanges filled single-particle levels with empty ones, a ground state invariant under $\Xi$ cannot be of Fermi-liquid type (at least not in the original electron degrees of freedom).

The theoretical treatment of the subject took off in 1993 with the work of Halperin--Lee--Read \cite{HLR}, who did propose a Fermi-liquid ground state for the lowest Landau level at half filling. Converting electrons into composite fermions \cite{Jain} by a procedure \cite{Girvin-MacD} called magnetic flux attachment, they argued that the latter could form a Fermi sea; the rough picture was that, by attaching two flux quanta to each electron, one cancels the background magnetic field on average, thus allowing the composite fermions to move as free fermions, at half filling. The technical step of flux attachment is carried out by introducing a fictitious gauge field, $a$, and adding to the field-theory Lagrangian a Chern-Simons term $a \wedge da$. (Actually, Read argued \cite{Read-CF1,Read-CF2} that the quantity bound to the electron should not be $\delta$-function Chern-Simons flux but rather vorticity.)

Although the HLR proposal was quite successful in fitting the observed phenomena, one bothersome issue remained: there exists no manifest particle-hole symmetry in the HLR field-theory Lagrangian. That's a serious worry because, as explained above, the Coulomb interaction projected to the lowest Landau level does have the particle-hole conjugation symmetry $\Xi$. Now much light and renewed interest has been thrown on the issue by a recent proposal of Son \cite{DTSon}, which we summarize briefly.

\subsection{Son's proposal}

Son \cite{DTSon} starts by observing that, for the purpose of developing a low-energy effective theory, one may realize the lowest Landau level as the subspace of zero modes of a massless Dirac fermion $\psi$, say with charge $q = +|e|$, in a homogeneous magnetic field:
\begin{equation}\label{eq:S-free}
    S = \mathrm{i}\hbar c \int dt \int d^2 r \, \bar\psi \, \gamma^\mu (\partial_\mu - \mathrm{i} q A_\mu / \hbar) \psi + ... \,,
\end{equation}
where the ellipses indicate residual interaction terms. In fact, adopting the symmetric gauge $A = |B| (x^1 dx^2 - x^2 dx^1) / 2$ for $B = |B| \, dx^1 \wedge dx^2$, and choosing the gamma matrices $\gamma^0 = \sigma_3\,$, $\gamma^1 = \mathrm{i} \sigma_2\,$, $\gamma^2 = - \mathrm{i} \sigma_1\,$, one arrives at a Dirac Hamiltonian $D$ of the form
\begin{equation*}
    D \propto \left( \begin{array}{cc} 0 & \partial_z - \bar{z}/4 \cr \partial_{\bar{z}} + z / 4 &0 \end{array} \right) , \quad z = (x^1 + \mathrm{i} x^2) \sqrt{|e B|/\hbar} \,,
\end{equation*}
and the zero modes of this Hamiltonian,
\begin{equation*}
    \psi_0 = \Bigg( \begin{array}{c} f(z) \cr 0 \end{array} \Bigg) \mathrm{e}^{-|z|^2 / 4} , \quad \partial_{\bar{z}} f(z) = 0 ,
\end{equation*}
are in bijection with the states spanning the lowest Landau level; cf.\ Section \ref{sect:Girvin}.

For the relativistic system (\ref{eq:S-free}), one has command of the discrete symmetry operations of charge conjugation $C$, parity $P$, and time reversal $T$. The product $CT$ is antilinear in second quantization and sends the electromagnetic field $(E,B)$ to $(-E,B)$. Thus it is an anti-unitary symmetry of the massless Dirac fermion (\ref{eq:S-free}) in zero electric field $E$ and for any magnetic field $B$. It is straightforward to check that $CT$ coincides with our operation $\Xi$ of particle-hole conjugation upon restriction to the zero-energy Landau level of the theory (\ref{eq:S-free}). Indeed, from Eqs.\ (\ref{eq:C-2plus1}, \ref{eq:T-2plus1}) we see that $CT$ acts in first quantization as the linear operator of multiplication by $\sigma_3\,$. This action anti-commutes with the Dirac magnetic Hamiltonian $D$ and is trivial on its zero modes (finite in number for a finite geometry). Returning to second quantization, we conclude that $CT \equiv \Xi$ on the (finite-dimensional) space of Dirac magnetic zero modes.

Let us emphasize once again that a $\Xi$-symmetric half-filled Fermi-liquid ground state does not exist, neither in the quantum Hall electron variables nor in the low-energy equivalent theory (\ref{eq:S-free}). In view of that no-go situation, one is motivated to look for a good change of variables by which to develop a Fermi-liquid description of some sort. (One such change of variables, of course, is the singular gauge transformation employed by HLR \cite{HLR} for non-relativistic electrons.)

Assuming the starting point (\ref{eq:S-free}), Son \cite{DTSon} performs a so-called fermionic particle-vortex transformation to pass to a dual formulation (known as ${\rm QED}_3$) by another Dirac field $\xi$ coupled to a dynamical gauge field $a$ (which coincides with the Chern-Simons dynamical gauge field $a$ of HLR but for a pseudoscalar multiplicative factor):
\begin{equation}\label{eq:S-dual}
    S_{\rm dual} = 2\pi\mathrm{i} \hbar v_F \int dt \int d^2 r \, \bar\xi \gamma^\mu (\partial_\mu - 2 \mathrm{i} a_\mu / q) \, \xi + \int A \wedge da + ... ,
\end{equation}
where we adopt the convention $dx^0 = v_F dt$ and $\partial_0 = v_F^{-1} \partial / \partial t$, as our physical system with characteristic speed $v_F$ has only Galilean invariance (not Lorentz invariance). This duality between ${\rm QED}_3$ and massless free fermions in $2+1$ dimensions has been much discussed and verified in the recent literature; cf.\ \cite{duality,duality4}.

Let us briefly touch on the physical content of the objects in the dual theory (\ref{eq:S-dual}). The dynamical gauge field $a = a_\mu dx^\mu$ is a gauge potential for the charge-current two-form $\rho - dt \wedge j \equiv J = da$ of the two-dimensional electron gas with electric charge density $\rho$ (more precisely, excess charge density $\rho \equiv \rho_{\rm excess}$ relative to half filling) and electric current density $j$. In particular, the time component $a_0$ is proportional to the orbital magnetization $m$ of the 2D electron gas. Please be advised that, recognizing $a$ as a quantity directly related to the physical observable $J$, we do \underline{not} speak of it as an ``emergent'' gauge field. Put differently, writing $J = da$ is simply a convenient way of taking care of the continuity equation $dJ = 0$ of charge conservation, in exactly the same way as writing $F = dA$ conveniently takes care of the homogeneous Maxwell equations $dF = 0$ (yet, nobody would say that the electromagnetic gauge field $A$ is ``emergent'').

The two-component spinor field $\xi$ in (\ref{eq:S-dual}) is called the Dirac composite fermion. It is charge-neutral, as it does not couple directly to the external gauge field $A$. In view of the basic duality between magnetic flux and electric charge, the coupling to the charge one-form $a$ suggests that $\xi$ carries an emergent magnetic flux. In fact, what $\xi$ carries is \emph{vorticity} (or pseudo-vorticity \cite{MS-RMP}), a quantity tied to the presence of magnetic flux.

The half-filled lowest Landau level features a nonzero orbital magnetization $\langle m \rangle$, and according to (\ref{eq:S-dual}) the magnetization $\langle m \rangle \sim \langle a_0 \rangle$ acts as a chemical potential for the Dirac composite fermion $\xi$. Therefore one may well expect the latter to form a Fermi-liquid ground state by populating a Fermi sea up to the chemical potential $\langle m \rangle$.

Let us finish here with the remark that the field theory (\ref{eq:S-dual}) contains neither a mass term for $\xi$ nor a Chern-Simons term for the gauge field $a$. In fact, both will turn out to be forbidden by the particle-hole conjugation symmetry $\Xi = CT$.

\subsection{Physical meaning of Son's theory}
\label{sect:deeper}

As a preparation for the microscopic picture of the composite fermion (Section \ref{sect:CF-micro}), we now delve somewhat deeper into the physical meaning of (\ref{eq:S-dual}). The present subsection is optional and may be skipped by readers interested only in the symmetry aspects of Son's effective field theory, which are elaborated in Sections \ref{sect:Sym-1st} and \ref{sect:Sym-2nd}.

\begin{figure}
    \begin{center}
        \epsfig{file=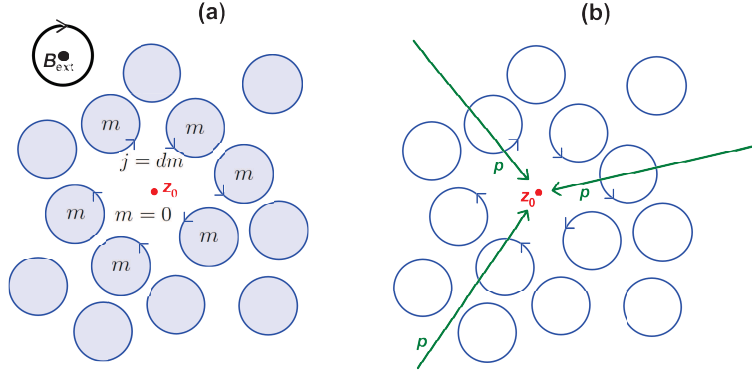,height=8cm}
    \caption{(a) Semiclassical picture of the magnetization $m$ and magnetization current $j = dm$ for the electron gas of a quantum Hall system (in an external magnetic field $B_{\rm ext}$) with a vortex centered at $z_0\,$. (b) Electric polarization $p$ associated with the vortex. The (gauge-dependent) one-form $p$ is visualized by (any) lines connecting the charge pushed to infinity with the charge deficit near $z_0\,$.}
    \label{fig:magpol}
    \end{center}
\end{figure}

Recall that the dynamical gauge field $a = a_\mu dx^\mu$ is a gauge potential for the charge-current two-form $J = \rho - dt \wedge j$. Writing $J = da$ and space-time decomposing
\begin{equation}
    a = m \, dt - p \,, \quad p = p_1\, dx^1 + p_2\, dx^2 ,
\end{equation}
one has $\rho = - {\rm d}p$ for the excess charge density and $j = {\rm d} m + \dot{p}$ for the electric current density, where ${\rm d}$ is understood to differentiate only with respect to space, not time. Using the traditional vector calculus of the physics literature, these expressions (written in three instead of two space dimensions) would take a form familiar from the textbook theory of electromagnetism in media:
\begin{equation*}
    \rho_{123} = - \mathrm{div}\, \vec{p} \,, \quad \vec{j} = \mathrm{rot}\, \vec{m} + \frac{\partial}{\partial t} \vec{p} \,.
\end{equation*}
After transcription from 3D to 2D, this analogy prompts us to interpret the expressions
\begin{equation}\label{eq:expr-pm}
    \rho = - {\rm d}p \,, \quad j = {\rm d} m + \dot{p} \,,
\end{equation}
or, in components,
\begin{equation*}
    \rho_{12} = - (\partial_1 p_2 - \partial_2 p_1), \quad j_l =
    \partial_l m + \frac{\partial}{\partial t} p_l \quad (l = 1, 2),
\end{equation*}
as saying that $p$ is an electric polarization one-form and $m$ an orbital magnetization function for the 2D electron gas (Fig.\ \ref{fig:magpol}). The latter are determined only up to gauge transformations
\begin{equation*}
    m \mapsto m + \frac{\partial \phi}{\partial t} \,, \quad
    p \mapsto p - {\rm d} \phi \,,
\end{equation*}
by a pseudoscalar function $\phi$ with the physical dimension of electric charge. Anticipating the discussion in Sect.\ \ref{sect:Sym-1st} below, we note that under time reversal one has $T : \; m \mapsto -m$ and $T :\; p \to +p$. Thus the gauge field $a = m\, dt - p$ is what is called a time-even form.

\medskip\noindent{\it Remark.} It is perhaps illuminating at this point to take a time-out and comment on the relation between our gauge field $a \equiv a_{\rm Son}$ and the Chern-Simons gauge field $a^\prime \equiv a_{\rm HLR}$ of Halperin--Lee--Read. Noting that the two differ in their total physical dimensions: $[a] = {\rm electric\, charge}$, versus $[a^\prime] = {\rm magnetic\, flux}$, one might expect them to be related by $(q / 4\pi\hbar)\, a^\prime = (1/q)\, a$. (In the present subsection, we continue to write $|e| \equiv q$, as we wish to reserve the symbol $e$ for an emergent electric field that will appear shortly.) Indeed, by the introduction of $a^\prime$ two magnetic flux quanta $4\pi\hbar/q$ are attached to each particle, while Son's gauge field $a$ simply does the tautological book-keeping job for the electric charge $q$ carried by each particle. Now electric charge and magnetic flux (and, correspondingly, $a$ and $a^\prime$) do not transform in the same way under orientation-reversing transformations of space-time. In fact, the charge-form $a$ is time-even and space-odd, whereas the flux-form $a^\prime$ is time-odd and space-even (see the beginning of Section \ref{sect:Sym-1st} for some detail). Thus to express one in terms of the other, a choice of space-time orientation $\mathrm{Or}_{2+1}$ needs to be brought into play \cite{footnote-OR}. The correctly stated relation is
\begin{equation}
    \frac{q^2}{4\pi\hbar}\, a_{\rm HLR} \cong a_{\rm Son} \otimes \mathrm{Or}_{2+1} \equiv (-a_{\rm Son}) \otimes (-\mathrm{Or}_{2+1}),
\end{equation}
or, in short: $a \cong \sigma_{\rm H} \, a^\prime$ with $| \sigma_{\rm H} | = q^2 /4\pi\hbar$ the Hall conductivity at half filling. (The symbol $\cong$ means equivalence up to gauge transformations: $\sigma_{\rm H} \, a^\prime = a + d\phi$.) A final remark is that Eqs.\ (\ref{eq:expr-pm}) turn into Eqs.\ (3) of Ref.\ \cite{Read-CF2} on substituting $a_{\rm Son} \to a_{\rm HLR}$. $\square$

Turning to the Dirac composite-fermion field $\xi$, we now expand on the statement that $\xi$ carries (pseudo-)vorticity by way of an emergent magnetic field. To start the argument, we observe that $S_{\rm dual}$ in (\ref{eq:S-dual}) has a symmetry under global $\mathrm{U}(1)$ phase rotations (not to be confused with gauge transformations $a \mapsto a + d \phi$ and $\xi \mapsto \mathrm{e}^{2\mathrm{i} \phi / q} \xi$),
\begin{equation*}
    \xi(x) \mapsto \mathrm{e}^{\mathrm{i}\theta} \xi(x) , \quad \bar\xi(x) \mapsto \mathrm{e}^{-\mathrm{i}\theta} \bar\xi(x) ,
\end{equation*}
which entails a conserved current:
\begin{equation*}
    \partial_\mu \Phi^\mu = 0 , \quad \Phi^\mu = \bar\xi \gamma^\mu \xi .
\end{equation*}
Hence, the physical meaning of $\xi$ hinges on the interpretation of the conservation law implied by $\partial_\mu \Phi^\mu = 0$. To uncover it, we introduce a two-form $f$ which is (Hodge-)dual to the given current vector field $\Phi = \Phi^\mu \partial_\mu :$
\begin{equation}\label{eq:CS-flux}
    f = \frac{1}{2} f_{\mu\nu} \, dx^\mu \wedge dx^\nu , \quad f_{\mu\nu}
    = (2h/q)\, \epsilon_{\mu\nu\lambda} \Phi^\lambda \quad (h \equiv
    2\pi\hbar),
\end{equation}
and we observe that by its construction from a conserved current, $f$ is closed ($d$$f = 0$). Now, integrating by parts ($\int A \wedge da = \int dA \wedge a$) and abbreviating the external electro\-magnetic field as $dA \equiv F$, we rewrite (\ref{eq:S-dual}) as
\begin{equation}\label{eq:S-MRZ}
    S_{\rm dual} = \mathrm{i} h v_F \int dt \int d^2 r \, \bar\xi
    \gamma^\mu \partial_\mu \xi + \int (f + F) \wedge a + ...
\end{equation}

Next, by the analogy with the electromagnetic field ($F = B + E \wedge dt$), we decompose the two-form $f$ as $f = b + e \wedge dt$, and we refer to $b$ and $e$ as emergent magnetic and electric fields. The conservation law due to the continuity equation $\partial_\mu \Phi^\mu = 0 = d$$f$ then reads $\int b = {\rm const}$ (independent of time). Moreover, taking the dual action (\ref{eq:S-MRZ}) for granted, we see that the functional integral over $a_0$ pins the emergent magnetic field $b$ to the external magnetic field $B$:
\begin{equation}\label{eq:pin-bB}
    B_{\rm eff} \equiv b + B = 0 .
\end{equation}
This constraint indicates that $\int b = {\rm const}$ reflects the conservation of magnetic flux ($\int B = {\rm const}$). Now, which conserved property of the electron gas is tied to the total external magnetic flux? There exists only one good answer to this question. To convey it, we note that any wave function for a particle of charge $q$ subject to a total magnetic flux of $\int B = N h/q$ has $N$ zeroes \cite{Poincare}. That basic property of single-particle wave functions carries over to the many-particle wave function, albeit in a somewhat opaque manner, as different particle coordinates may see some of the zeroes (namely, the ``non-Pauli'' zeroes \cite{MS-RMP}) in different places. Notwithstanding these many-body complications, an informative quantity to consider is the \emph{one-body density matrix} $\rho({\bf r}^\prime ,{\bf r})$ for the many-particle wave function, say, in a time-dependent situation with quasi-particle excitations in motion. (In fact, Son's proposal tries hard and does its very best to serve as a one-body or mean-field description!) Viewed as a function of one position variable, ${\bf r}$, that object does have $N$ well-defined and physically meaningful zeroes, whose locations vary with the other position variable, ${\bf r}^\prime$. We may call such a zero a \emph{vortex}, as the phase of the one-body density matrix increases by $2\pi$ along closed loops encircling it.

\begin{figure}
    \begin{center}
        \epsfig{file=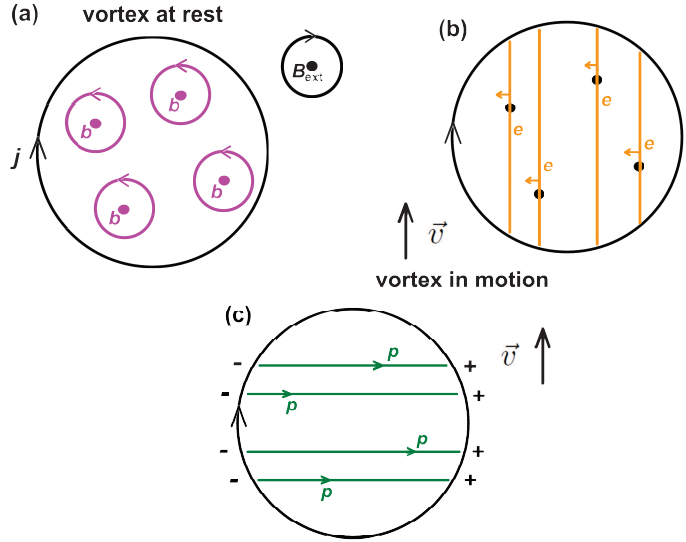,height=8cm}
    \caption{(a) Vortex at rest. (Secretly, what is shown here is a charge-neutral double vortex with an electron bound to it.) The vortex carries an emergent magnetic field $b$ (indicated by a few of its points) with circulation opposite to that of $B_{\rm ext}$. (b) In a state of motion with velocity $\vec{v}$, the vortex carries an emergent electric induction field $e = \iota(\vec{v}) b$. As a one-form, $e$ is visualized by its null lines with outer orientation by a transverse vector. (c) By the dielectric response of the electron gas, the effective electric field $E_{\rm eff} = e + E$ translates into an electric polarization field $p$. The corresponding charge density $\rho = - \mathrm{d}p$ for the vortex in motion is that of an electric dipole.} \label{fig:vortexmoves}
    \end{center}
\end{figure}

Adopting that terminology, we can say that the total number $N$ of vortices is fixed by the total magnetic flux as $N = (q/h) \int B$. In view of Eq.\ (\ref{eq:pin-bB}), it is then natural to interpret the emergent magnetic field $b$ (normalized by multiplication with $q/h$) as the ``density'' of vortices (cf.\ Fig.\ \ref{fig:vortexmoves}a). To further that interpretation, we observe that the decomposition $f = b + e \wedge dt$ turns the conservation law $d$$f = 0$ into Faraday's law of induction:
\begin{equation}
    \dot{b} + {\rm d} e = 0 ,
\end{equation}
whose physical content is that vorticity in motion ($\dot{b} \not= 0$) is accompanied by an emergent electric induction field $e$ (cf.\ Fig.\ \ref{fig:vortexmoves}b). A second equation involving $e$ is obtained as the equation of motion for $J = da$ that follows from (\ref{eq:S-dual}) by variation of the spatial components of $a$. To get a sensible answer, we should add to $S_{\rm dual}$ a dynamical term for $J = \rho - dt \wedge j$. Here the usual Maxwell term of ${\rm QED}_3\,$,
\begin{equation*}
    da \wedge \star \, da = \frac{1}{2} J_{\mu\nu} J^{\mu\nu} |d^3 x|\,,
    \quad J_{\mu\nu} = \partial_\mu a_\nu - \partial_\nu a_\mu \,,
\end{equation*}
is inappropriate, as our non-relativistic system does not obey Lorentz invariance and electric charges interact non-locally via electromagnetic fields in the ambient 3D space. Thus a more plausible addition to (\ref{eq:S-MRZ}) would be a charge-charge interaction of 3D-Coulomb type or similar. However, the simplest modification we can make is to add the interaction $\int V \rho$ with an electrostatic disorder potential $V$.
The resulting equation is
\begin{equation}
    E_{\rm eff} \equiv e + E = - \mathrm{d} V .
\end{equation}
What enters here on the left-hand side is an effective field strength $E + e$, gotten by adding to the external electric field $E$ the emergent electric field $e$. The micro\-scopic interpretation is that the composite fermion carries an electric dipole moment (Fig. \ref{fig:vortexmoves}c).

We have yet to comment on the expression $\nabla_\mu = \partial_\mu - 2 \mathrm{i}\,a_\mu / q$ for the covariant derivative in (\ref{eq:S-dual}). If the space manifold $M$ is taken to be compact and the charge density does not integrate to zero ($\int_M \rho = \int_M J \not= 0$), any gauge potential $p = - a_l \, dx^l$ for $\rho$ exists only locally and we are dealing with a nontrivial line bundle (twisting the Dirac spinor bundle). Its covariant derivative has curvature
\begin{equation*}
    [ \nabla_\mu \,, \nabla_\nu ] = - 2 \mathrm{i} J_{\mu\nu} / q \,,
\end{equation*}
or $\nabla^2 = - 2\mathrm{i} J / q$. The Dirac quantization condition then forces the allowed charge totals $Q = \int_M \rho$ to be in the lattice $Q \in q \mathbb{Z}/2$ \cite{foot:anomal}. Correspondingly, canonical quantization of the theory (\ref{eq:S-dual}) will make for single-particle excitations of $\xi$ that carry \emph{two} emergent magnetic flux quanta $2h/q$ and hence two vortices due to the constraint (\ref{eq:pin-bB}).

To finish this excursion on a philosophical note, it has long been appreciated that the LLL composite fermion is a composite object made from one electric charge quantum (electron) and two magnetic flux quanta (vortices). The traditional approach of HLR was to build the theory around the electron degrees of freedom and attach Chern-Simons magnetic flux as a subsidiary feature. The recent approach of Son turned the emphasis around, by taking the vortex degrees of freedom (which carry magnetic flux but no electric charge) as primary and coupling them to fluctuations in the charge current $J$ via the magnetization/polarization one-form $a$. While the physics should ultimately come out to be the same \cite{WCHS-PRX}, the computational details differ between the two approaches. For one major difference, the number of composite fermions in the HLR approach is given by the number of electrons, whereas in Son's proposal it is given by half the number of magnetic flux quanta. (Note that the two definitions coincide at half filling but differ away from it.) Another major difference, namely the realization of symmetries, will be our focus in the next two subsections.

\subsection{Symmetry considerations: first quantization}
\label{sect:Sym-1st}

Our main interest here is in the symmetry aspects of Son's proposal and especially in the issue of particle-hole (conjugation) symmetry. Since particle-hole conjugation $\Xi$ is realized in the (zero Landau level of the) Dirac fermion representation (\ref{eq:S-free}) by $CT$, we shall now elucidate the corresponding operation $CT$ in the dual representation (\ref{eq:S-dual}).

As we will see, the operation of time reversal $T$ is realized on the fermionic vortex field $\xi$ in an unfamiliar way making the symmetry aspects quite striking. To spell them out, we begin by reviewing how $T$ acts on the electromagnetic gauge field $A$ and on the charge current $J$. Our guiding principle here is that the field-matter interaction $A \wedge J$ must transform as a space-time density, so that
\begin{equation*}
    S_{\rm int} = \int A \wedge J = \frac{1}{2} \int A_\mu dx^\mu \wedge J_{\nu \lambda}\, dx^\nu \wedge dx^\lambda = \int A_\mu J^{\mu} | d^3 x |
\end{equation*}
is invariant under all space-time diffeomorphisms including those that are orientation reversing, and is invariant under time reversal in particular. Now the two-form $J = da$ for the charge 3-current (in $2 + 1$ dimensions) is a time-even differential form, which means that it transforms under time reversal $T$ by straight pullback: $J \mapsto + T^\ast J$. It then follows from $J = da$ and $T^\ast \circ d = d \circ T^\ast$ that the magnetization/polarization one-form $a = a_\mu d x^\mu$ is also time-even: $a \mapsto + T^\ast a$; in components we have
\begin{equation*}
    T :\; a_0 ({\bf r},t) \mapsto - a_0({\bf r},-t) , \quad
    a_l ({\bf r},t) \mapsto + a_l({\bf r},-t) \quad (l = 1, 2).
\end{equation*}
On the other hand, the electromagnetic gauge field $A$ is a time-odd one-form; thus it transforms under time reversal by negative pullback ($A \mapsto - T^\ast A$) or
\begin{equation*}
    T :\; A_0 ({\bf r},t) \mapsto + A_0({\bf r},-t) , \quad A_l ({\bf r},t) \mapsto - A_l({\bf r},-t) \quad (l = 1, 2).
\end{equation*}
(Of course, the opposite behavior of $J \mapsto + T^\ast J$ versus $A \mapsto - T^\ast A$ is just what is needed in order for $A \wedge J = A_\mu J^{\mu} | d^3 x |$ to transform as a scalar under time reversal.)

The sign-opposite transformation law for $a$ as compared with $A$ has a surprising effect. To see that most clearly, consider the first-quantized Hamiltonian $h$ of the vortex field $\xi$ in a given gauge field $a = m \, dt - p :$
\begin{equation}\label{eq:h-1st}
    h(p,m) = \hbar v_F \sum_{l=1}^2 \sigma_l \left( \frac{1} {\mathrm{i}} \frac{\partial}{\partial x^l} + \frac{2}{q} \, p_l \right) - \frac{2\hbar}{q} \, m \,,
\end{equation}
where $m = v_F a_0$ is the local magnetization, and $p = \sum p_l \, dx^l$ is the local polarization. To determine how $T$ acts on $\xi$, one observes that if $\xi$ is a solution of the Dirac equation $\mathrm{i} \hbar \partial_t \xi = h \xi$, then by time-reversal invariance (or equivariance) so is $T \xi$. There exist two different scenarios by which to realize that equivariance condition. The standard scenario is that $T$ commutes with both $h$ and $\mathrm{i} \hbar \partial_t\,$. Since $\partial_t$ changes sign under $t \mapsto -t$, this means that $T$ must be antilinear ($T \mathrm{i} = - \mathrm{i}\, T$). The second scenario for time-reversal symmetry of the Dirac equation is that $T$ \emph{anti}-commutes with both $h$ and $\mathrm{i} \hbar \partial_t\,$. (Recall that we have already met this scenario for the case of a particle-hole symmetry $K$, in Section \ref{sect:path-int} on path-integral realization. The possibility for such a scenario to exist has been strongly emphasized in \cite{FM2013}.) Given Eq.\ (\ref{eq:h-1st}), it is clear that the time-even property of $a$, namely $T : \; m \mapsto - m$ and $p_l \mapsto + p_l$ ($l = 1, 2$), forces the latter scenario. Thus in the present instance, time reversal is realized as an operation that is complex \emph{linear} ($T \mathrm{i} = + \mathrm{i}\, T$) and anti-commutes with the vortex-field Hamiltonian. Explicitly,
\begin{equation}
    T : \; \xi({\bf r},t) \mapsto \sigma_3 \xi({\bf r},-t) , \quad h(p,m) \mapsto \sigma_3 h(p,-m) \sigma_3 = - h(p,m).
\end{equation}
This concludes our discussion (in first quantization) of how time reversal is realized on the fermionic vortex field $\xi$ of the dual representation.

The other factor of the symmetry operation $CT$ is charge conjugation $C$. We recall that the electromagnetic gauge field $A$ transforms under $C$ as $A \mapsto - A$. In order for the field-matter interaction $\int A \wedge da$ to be charge-conjugation invariant, the dynamical gauge field $a$ must conform to the same transformation law as $A$ (thus, $C : \; a \mapsto -a$). By this coincidence, we may take the actions of $C$ on the fermionic vortex field $\xi$ and the Dirac field $\psi$ to be identical. From Eq.\ (\ref{eq:C-2plus1}) we then have
\begin{equation*}
    C : \; \xi \mapsto \sigma_1 \bar{\xi} .
\end{equation*}
(Here we abandon the pedagogical chore of using different symbols for charge conjugation in first and second quantization.) Thus $C$ is antilinear in first quantization and anti-commutes with the operator $\mathrm{i} \hbar \partial_t - h$:
\begin{equation*}
    C :\; h(p,m) \mapsto \sigma_1 \overline{h(-p,-m)} \sigma_1 = -h(p,m).
\end{equation*}

Next, we turn to the combined operation $CT$. By the properties of its factors $C$ and $T$, the product $CT$ is an antilinear symmetry of the first-quantized Hamiltonian:
\begin{equation}
    CT : \; h(p,m) \mapsto \sigma_2 \, \overline{h(-p,m)} \sigma_2 = + h(p,m) .
\end{equation}
Note that our emphasis here is on \emph{symmetry} (as opposed to anti-symmetry $h \mapsto -h$).

The field-matter interaction $\int A \wedge J$ for $J = da$,
\begin{equation*}
    \int A \wedge da = \int dA \wedge a = \int (B + E \wedge dt) \wedge (m\, dt - p) ,
\end{equation*}
augments the vortex-field Hamiltonian by a term
\begin{equation*}
    h_{\rm int} = \int (B m - E \wedge p) ,
\end{equation*}
which is invariant under both $T$ and $C$. The combined action of these operations on the electro\-magnetic field and the dynamical gauge field is
\begin{equation*}
    (E,B) \stackrel{CT}{\mapsto} (-E,B), \quad (p,m) \stackrel{CT}{\mapsto} (-p,m) .
\end{equation*}
Now from our condensed-matter perspective, the electromagnetic field is to be regarded as a given background (not to be transformed). We then see that $CT$ remains a symmetry for $E \equiv 0$ (zero external electric field) and any magnetic field $B$.

We are finally ready to deliver the symmetry punch line of Son's proposal. In the original formulation (\ref{eq:S-free}) by a massless Dirac fermion projected to the zero Landau level, the operator $CT$ acted as particle-hole conjugation $\Xi$. Thus it exchanged particles and holes (or antiparticles) in second quantization, thereby posing an obstruction to the existence of any Fermi-liquid ground state (with symmetry $\Xi$). Now in the dual re\-presentation (\ref{eq:S-dual}) this obstruction has disappeared! Indeed, the operation $CT$ on the fermionic vortex field $\xi$ and the dynamical gauge field $a$ is a proper symmetry of the Hamiltonian; thus it sends particles (actually, particle-like excitations of the vortex field) to particles and antiparticles to antiparticles. In Son's language, the Dirac composite fermion $\xi$ is its own antiparticle. This, then, is how $CT$ (alias particle-hole conjugation) may emerge as a symmetry of a Fermi-liquid ground state.

\subsection{Symmetry considerations: second quantization}
\label{sect:Sym-2nd}

Up to now, we have discussed the dual representation (\ref{eq:S-dual}) in the formalism of first quantization. Of course, the same conclusions are reached by considering the second-quantized theory as in \cite{DTSon}. Let us briefly indicate the main aspects.

First of all, the unfamiliar situation with the action of time reversal $T$ gets rectified by passing to second quantization with Hamiltonian $H = \int d^2 r \, \xi^\dagger h \xi$. Because $T$ anti-commutes with $h$, it gets concatenated with the antilinear operation $\xi \leftrightarrow \xi^\dagger$ of particle-hole conjugation so as to become a symmetry of the vortex-field Hamiltonian $H$:
\begin{equation*}
    T : \; H(p,m) \mapsto \int d^2r \, \xi \sigma_3 \overline{h(p,-m)} \sigma_3 \xi^\dagger = + H(p,m) \quad (B = 0).
\end{equation*}
Thus, in second quantization we are back to the standard realization of time reversal $T$ as an antilinear operation that commutes with the Hamiltonian. Nonetheless, one non-standard feature does remain: the second-quantized $T$ involves the operation of particle-hole conjugation! As before, it follows that the $CT$-operation
\begin{equation}\label{eq:CT-dual}
    CT : \; H(p,m) \mapsto \int d^2r \, \xi^\dagger \sigma_2 \overline{h(-p,m)} \sigma_2 \xi = + H(p,m)
\end{equation}
is antilinear, while sending $\xi$ to $\xi$ and $\xi^\dagger$ to $\xi^\dagger$ (due to two particle-hole conjugations, one from $C$ and another one from $T$), which is consistent with the hypothesis of a Fermi-liquid ground state for the Dirac composite fermion.

Secondly, we add for completeness that the situation with parity $P$ (meaning any orientation-reversing reflection of space) is analogous to that for $T$. In first quantization, $P$ acting on $\xi$ is antilinear and anti-commutes with the Hamiltonian. However, in second quantization we go back to the standard situation: $P$ is then linear and commutes with the Hamiltonian (if $B = 0$). As with $T$, the unusual feature with the second-quantized $P$ is that it involves the operation of particle-hole conjugation. Thus, all of the three operations $C$, $P$, and $T$ on the fermionic vortex field $\xi$ contain particle-hole conjugation and thus exchange particles for antiparticles -- a point stressed by Son \cite{DTSon}.

Thirdly, it should be emphasized that the $CT$-symmetry forbids a mass term for $\xi$ and a Chern-Simons term for $a$. Indeed, the CS-action $\int a \wedge da$ needs an orientation of space-time and changes sign when that orientation is reversed (as happens under $CT$). Similarly, a mass term $H_{\rm mass} \propto \int d^2r\, \xi^\dagger \sigma_3 \xi$ changes sign under $CT$; see (\ref{eq:CT-dual}).

For a final statement, one might say \cite{duality4} that $CT$ and $T$ get exchanged under fermionic particle-vortex duality in $2+1$ dimensions. While that is a sensible and informative viewpoint, we prefer to say that the well-defined operation of time reversal $T$ remains $T$, and likewise, $CT$ remains $CT$. The truth is simply that the spinor fields $\psi$ and $\xi$ on the two sides of the duality are distinct physical objects with \emph{different} behavior under space-time transformations, as $\psi$ carries electric charge whereas $\xi$ carries emergent magnetic flux, and therefore, the time-reversal operation $T$ must act on $\psi$ in one way and on $\xi$ in another way (actually, as $CT$), and the same goes for $CT$.

\subsection{Comments: microscopic picture of the composite fermion}
\label{sect:CF-micro}

Since the field-theoretic symmetry considerations may seem quite abstract, one might be inclined to look into the microscopic picture of the composite fermion. Not being an expert, the author is somewhat hesitant to do so, as the theory of the half-filled Landau level remains controversial in spite of its long road of development (a few references are \cite{Read-nu1,MS-RMP,DCF-WS,GZMMVM,MS2015,WCHS-PRX}), and the same goes for the microscopic picture. Nonetheless, a few remarks can and should be made from the present vantage point.

For quantum Hall states and other systems with an energy gap where the quantum adiabatic theorem holds, one knows \cite{ASS-PRL} or expects \cite{Laughlin,MS-RMP} that the adiabatic insertion of a magnetic flux line (in 3D, or flux point in 2D), with circulation equal to that of the strong magnetic background field, expands the electron gas radially outward from the point of insertion. The adiabatic flux insertion gives rise to spectral flow resulting in a zero (or vortex) of the many-electron wave function. Adopting the symmetric gauge with respect to the insertion point, $z_0\,$, one can express the effect of flux insertion as multiplication by the operator
\begin{equation}\label{eq:vortex-op}
    U_{z_0} = \prod\nolimits_j (z_j - z_0) ,
\end{equation}
where $z_j$ ($j = 1, 2, ...$) are the electron coordinates of the holomorphic representation (see Sect.\ \ref{sect:Girvin}). In the occupation-number representation ${\bf n} = \{ n_0, n_1, n_2 , ... \}$ w.r.t.\ the single-particle basis $(z-z_0)^l \mathrm{e}^{-|z- z_0|^2 / 4}$ ($l = 0, 1, 2, ...$), the effect of the vortex operator $U_{z_0}$ is a shift ${\bf n} \mapsto \{ 0, n_0, n_1, \ldots \}$ leaving the zero orbital ($l = 0$) vacant. The vacancy amounts to a local charge deficit: for filling fraction $\nu$, the charge of the vortex $-\nu e$.

For gapless systems such as the half-filled lowest Landau level ($\nu = 1/2$), the protocol of adiabatic flux insertion rests on less firm grounds. However, it still seems reasonable to assume that multiplication by the vortex operator (\ref{eq:vortex-op}) creates a low-energy excitation of the electron gas, with charge $- \nu e = - e/2$ in our case.

Now if $U_{z_0}$ (or rather, its plane-wave descendants near a Fermi surface) plays a role in the low-energy effective theory, then by the particle-hole conjugation symmetry of the half-filled lowest Landau level, one would expect the particle-hole conjugate
\begin{equation}\label{eq:antivortex}
    U_{z_0}^\flat = \Xi U_{z_0} \Xi^{-1}
\end{equation}
to play an equally important role. So, what is $U_{z_0}^\flat$? The answer is found most easily in the occupation-number representation of above. Recalling from Sections \ref{sect:lifting-PH} and \ref{sect:phsym-gapless} that $\Xi$ simply conjugates the occupation numbers ($n_l \mapsto 1-n_l$) while possibly changing the overall sign of the wave function, we see immediately that $U_{z_0}^\flat$ has exactly the same effect as $U_{z_0}$ except for one characteristic difference: while $U_{z_0}$ leaves the zero orbital always empty, the particle-hole conjugate $U_{z_0}^\flat$ leaves it always occupied. Thus $U_{z_0}^\flat$ creates a vortex centered at $z_0$ and at the same time adds an electron in the zero orbital at $z_0\,$. In view of the genesis (\ref{eq:antivortex}) we will often refer to  $U_{z_0}^\flat$ as an ``anti-vortex'' operator.

Guided by the traditional idea that composite fermions carry an \emph{even} number of magnetic flux quanta (here, two), let us now consider the vortex--anti-vortex composites
\begin{equation*}
    C_{z_0} = U_{z_0}^\flat U_{z_0}^{\vphantom{\flat}} \,, \quad
    C_{z_0}^\flat = \Xi C_{z_0} \Xi^{-1} = \pm   U_{z_0}^{\vphantom{\flat}} U_{z_0}^\flat \,,
\end{equation*}
which create a single electron along with a double vortex, i.e.\ a zero of order two in the many-electron wave function, at $z_0\,$. (The sign due to $U_{z_0}^{\flat\flat} = \pm U_{z_0}$ alternates with the Landau level dimension $N$.) Both are maps
\begin{equation*}
    C_{z_0} \,,\, C_{z_0}^\flat :\quad {\textstyle{\bigwedge}}^n (\mathbb{C}^N) \to {\textstyle{\bigwedge}}^{n+1}(\mathbb{C}^{N+2})
\end{equation*}
from the Fock space for the lowest Landau level with $N$ magnetic flux quanta to that with $N+2$ flux quanta. In particular, they both send a state at half filling, i.e.\ $n/N = 1/2$, to another state at half filling: $(n+1)/(N+2) = 1/2$. In our occupation-number representation they act as
\begin{equation}\label{eq:C-occno}
    \begin{array}{ll}
    \pm C_{z_0} : \; (n_0, n_1, n_2, ..., n_{N-1}) \mapsto (1, 0, n_0,
    n_1, n_2, ..., n_{N-1}) , \cr \pm C_{z_0}^\flat : \; (n_0, n_1, n_2,
    ..., n_{N-1}) \mapsto (0, 1, n_0, n_1, n_2, ..., n_{N-1}) .
    \end{array}
\end{equation}
Note especially that $C_{z_0}$ and $C_{z_0}^\flat$ are defined as particle-hole conjugates of each other. They constitute a ``Kramers pair'' (w.r.t.\ the antilinear symmetry $\Xi$) in the sense that
\begin{equation}\label{eq:2flat}
    C_{z_0}^{\flat\flat} = \Xi^2 C_{z_0} \Xi^{-2} = (-1)^{(N+2)(N+1)/2} C_{z_0} (-1)^{N(N-1)/2} = - C_{z_0}
\end{equation}
due to Eq.\ (\ref{eq:4.10-new}). Another joint feature is that they create local excitations which are electrically neutral. Indeed, the presence of the double vortex implies a local charge deficit of $2 \times (-\nu e) = -e$ at half filling, and this is exactly compensated by the addition of a single electron with charge $e$. The charge neutrality at $\nu = 1/2$ is also clear from the expressions (\ref{eq:C-occno}) in the occupation-number representation.

So, the assumed principle of particle-hole conjugation symmetry hands us a \emph{pair} of conjugate operators, $C_{z_0}$ and $C_{z_0}^\flat$ (for any $z_0$). To put this message into context, recall that the creation operator for a bare electron $\Xi$-conjugates to an annihilation operator (leading to the conundrum of how to form a particle-hole conjugation symmetric ground state of Fermi-liquid type). In contrast, when the particle-hole conjugation operator $\Xi$ is applied to the composite object $C_{z_0}$ creating an electron along with a double vortex, we obtain another object $C_{z_0}^\flat$ of the very same kind -- the only difference is that $C_{z_0}$ puts the electron into the $l=0$ orbital at $z_0\,$, whereas $C_{z_0}^\flat$ puts it into the $l=1$ orbital.

Can we identify the duplicity between $C_{z_0}$ and $C_{z_0}^\flat$ with the structure of Son's Dirac composite fermion as a two-component spinor? Alas, the answer is no --- our arguments are beset with a fallacy! The problem is that by letting our putative composite-fermion creation operators $C_{z_0}$ and $C_{z_0}^\flat$ act on a homogeneous fluid, we arrive at two different charge distributions, of which only the former one ($l = 0$) minimizes the Coulomb interaction energy with the surrounding electron gas. Therefore, our statement that $C_{z_0}$ and $C_{z_0}^\flat$ constitute a symmetry-related pair appears to be false.

What's the error in the reasoning above? Is the $\Xi$-symmetry spontaneously broken \cite{BMF-2015}? Or is the comparison between $C_{z_0}$ and $C_{z_0}^\flat$ better made for the zero-momentum operators constructed from them, both of which are expected to annihilate the Fermi-sea ground state of composite fermions (there is no contradiction in $0 = 0$)? Or does there exist another, more attractive alternative to be explored?

Let's get some perspective. Hanging on to old ideas, we have been assuming the picture of composite fermions formed by attaching two $\delta$-function fluxes to an electron or, in a slight improvement, placing the vortex ($U_{z_0}$) of charge $-e/2$ and the anti-vortex ($U_{z_0^\prime}^\flat$) of charge $+e/2$ right on top of each other ($z_0 = z_0^\prime$). Yet, there is no compelling reason (other than simplicity) to make that assumption, and the vortex--anti-vortex pair at rest might rather form a more or less loosely bound electric dipole; if so, the contradiction due to the fake Kramers pair $C_{z_0}$ with $C_{z_0}^\flat$ goes away. Along these lines, Wang--Senthil \cite{DCF-WS} have suggested the following intuitive picture.

\begin{figure}
    \begin{center}
        \epsfig{file=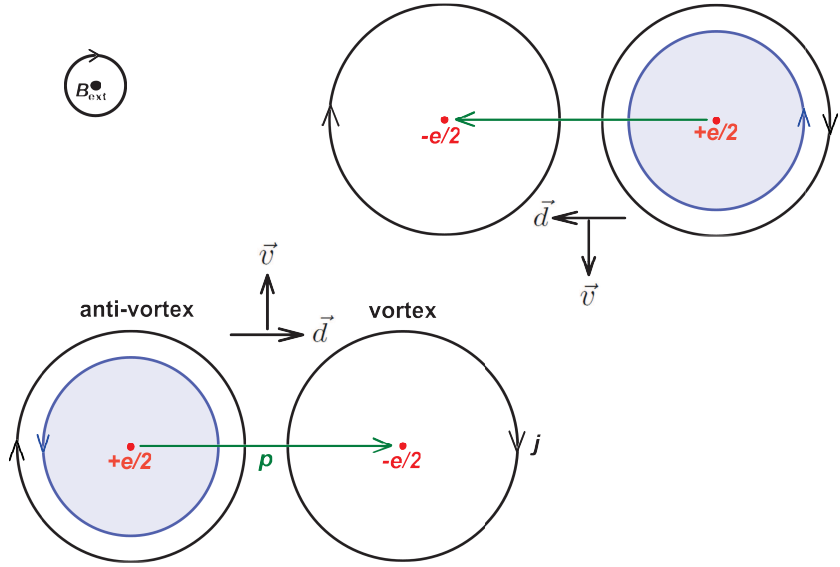,height=8cm}
    \caption{Semiclassical picture at $\nu = 1/2$ of a composite fermion (left-hand side) and its particle-hole conjugate (right-hand side). The composite fermion is an electric dipole made from a vortex with positive charge $-e/2$ and an anti-vortex with negative charge $+e/2$. In a state of motion, the velocity $\vec{v}$ and the dipole moment $\vec{d}$ form an orthogonal pair positively oriented with respect to $B_{\rm ext}\,$.}
    \label{fig:dcf}
    \end{center}
\end{figure}

One thinks of the many-electron wave function for the half-filled lowest Landau level as containing vortices with charge quanta $\pm e/2$. The two vortex types are particle-hole conjugates of each other. A repulsive (resp.\ attractive) electric force acts between vortices of the same (resp.\ opposite) type. The proposed picture is that low-energy excitations (of the ground state at half filling) are fast moving dipoles made from a vortex bound to an anti-vortex (Fig.\ \ref{fig:dcf}). What remains unclear to the present author is whether these composite fermions exist as such over the wide range of wave numbers $|k| \leq k_F$ of a Fermi sea. In fact, a pessimist might say that the quantitative verification of the Fermi-liquid scenario appears rather challenging, as there exists no obvious parameter to expand in --- we expect the size of the composite fermion to be comparable to the vortex size, which in turn is of the order of the mean distance between the vortices.

Nevertheless, to add a bit of quantitative detail, Wang--Senthil (\cite{DCF-WS}, WS) have considered the extreme limit of composite fermions made from vortex--anti-vortex pairs with mutual separation much larger than the vortex size. In that limit, the self-statistics and mutual statistics of the vortex and anti-vortex constituents is well-defined and can be used for an easy demonstration that the Berry phase for the adiabatic process of dragging one around the other is $\pi$. Further, in a much simplified model (replacing the pair of vortices by a pair of point charges in a strong magnetic field) WS demonstrate the property of ``spin-momentum locking'': the velocity of motion of the composite fermion is perpendicular to the vector of its electric dipole moment. It then follows from the corresponding calculation for massless Dirac fermions that the process of adiabatically transporting a composite fermion around the Fermi surface, or rotating its $k$-vector through $2\pi$, gives a Berry phase of $\pi$. (The latter had been recognized as an important ingredient \cite{FDMH-Berry,DTSon} for the correct phenomenology of the composite Fermi liquid.) The caveat with this line of arguments is the questionable assumption of a clear separation of length scales (namely, composite fermion size much larger than vortex size).

On that cautious note, we end our account of the half-filled lowest Landau level and its composite fermion. We have included it here for the starring role that particle-hole conjugation symmetry plays in it.

\section{Summary}\label{sect:conclude}

Let us summarize the key messages of the paper.

\begin{itemize}
\item Fundamental to the notion of particle-hole symmetries is the tautological operation of particle-hole conjugation. On operators $L$ acting on quantum states, the latter is defined as the algebra automorphism $L \mapsto L^\flat$ which exchanges single-particle creation and annihilation operators ($a^\dagger \leftrightarrow a$) while keeping the operator order fixed: $(AB)^\flat = A^\flat B^\flat$ (recall that Hermitian conjugation $L \mapsto L^\dagger$ reverses the operator order). Particle-hole conjugation shares with Hermitian conjugation the property of being antilinear: $(z L)^\flat = \bar{z} L^\flat$ ($z \in \mathbb{C}$). That property is forced, since single-particle creation and annihilation operators are modeled after the single-particle Hilbert space $V$ and its dual $V^\ast$, respectively, and there exists no invariant way of identifying $V$ with $V^\ast$ without employing the Dirac ket-to-bra bijection (or Fr\'echet-Riesz isomorphism) $V \ni \vert v \rangle \mapsto \langle v \vert \in V^\ast$, which is antilinear.
\item For a single-particle Hilbert space $V_0$ of dimension $N < \infty$, particle-hole con\-jugation $L \mapsto L^\flat$ lifts as $L^\flat = \Xi L \Xi^{-1}$ to an operation $\Xi : \; \bigwedge^n(V_0) \to \bigwedge^{N-n}(V_0)$ on the fermionic Fock space. A basis-free construction of that lift was given in Section \ref{sect:lifting-PH}, by concatenating the Fr\'echet-Riesz isomorphism $\bigwedge^n(V_0) \to \bigwedge^n(V_0)^\ast$ with a wedge isomorphism $\bigwedge^n(V_0)^\ast \to \bigwedge^{N-n} (V_0)$. As an important corollary, the antilinear operator $\Xi$ has square $\Xi^2 = (-1)^{N(N-1)/2}\, \mathbf{1}_{\wedge(V_0)}$ (dependent on the Hilbert space dimension $N$ but not on the particle number $n$).
\item Particle-hole conjugation would be an antilinear symmetry of the many-body Hamiltonian $H$ if $H^\flat = H$. Yet, any Hermitian and traceless one-body Hamiltonian $H$ is particle-hole odd: $H^\flat = - H$. It follows that particle-hole conjugation cannot ever be a symmetry of any system close to free fermions. (However, one may have $H^\flat = H$ in situations like that of the lowest Landau level, where the kinetic energy is quenched.) We emphasize that the relation $H^\flat = - H$ also holds for the Hartree-Fock-Bogoliubov mean-field Hamiltonian $H$ of any superconductor or superfluid and is no more than an algebraic tautology (due to the canonical anti-commutation relations for fermions). It should not be mistaken for a ``particle-hole symmetry''!
\item In order for a system of free fermions to have a particle-hole symmetry, $K$, some additional structure is needed (to undo the sign reversal $H^\flat = - H$). That structure is an involution $\Gamma$ (initially defined on the single-particle Hilbert space, and then extended to the Fock space), which sign-reverses the first-quantized Hamiltonian (or the BdG-Hamiltonian in the case of pair-condensed systems treated in mean-field approximation) and thus exchanges the single-particle states of positive energy with those of negative energy. Although $\Gamma$ might in principle be antilinear (akin to the charge-conjugation operation of the first-quantized Dirac theory), it is linear in all the examples we know from condensed matter physics -- typically, $\Gamma$ derives from a sublattice transformation or momentum shift. A particle-hole symmetry $K$ is then given as $K = \Xi \circ \Gamma$, with $\Xi$ replaced by a suitable substitute if $N = \infty$. To promote the $K$-symmetry of the Hamiltonian to a symmetry of the free-fermion ground state, the system must be at half filling or zero chemical potential.
\item As was demonstrated with a number of examples, our notion of particle-hole symmetry continues to make sense in the realm of interacting systems. For one major application, we argued that the Haldane phase of antiferromagnetic spin chains is adiabatically connected to an $\AIII$-symmetry protected free-fermion topological phase. For another, we gave a detailed symmetry analysis of Son's proposal for a particle-hole conjugation symmetric effective field theory of the half-filled lowest Landau level. We were also able to throw some light on the debate concerning the microscopic picture of the composite fermion.
\item An illuminating message delivered by the example of the half-filled lowest Landau level is this: under fermionic particle-vortex duality in $2+1$ dimensions, the standard time-reversal symmetry $T$ on the particle side turns into a symmetry of particle-hole type (alias $K$) on the vortex side. This correspondence under duality suggests that, more  generally, $T$ and $K$ are close cousins as antilinear transformations inverting the time direction --- a point already made in Section \ref{sect:path-int} using the language of the path integral. A characteristic difference between $T$ and $K$ lies in how they act on the electric charge ($Q$) and the electric current ($I$); one has $T : (Q,I) \mapsto (+Q,-I)$ as opposed to $K :\; (Q,I) \mapsto (-Q,+I)$. By the standard coupling to the electromagnetic field, it follows that $T$ as a symmetry tolerates the presence of an external electric field, whereas $K$ tolerates an external magnetic field. In this sense, $T$ and $K$ are complementary: both are needed to encompass the plethora of phenomenology encountered in condensed-matter systems.
\item Our definition of particle-hole symmetry $K$ as an anti-unitary operation agrees with the one given in the original paper \cite{HHZ} stating and proving the Tenfold Way classification (see also our follow-ups \cite{OHRMT,KZ}), and differs from that invented in \cite{SRFL2010,CTSR2016} and called the ``AZ classification'' in reference to \cite{AZ97}. In fact, \cite{SRFL2010,CTSR2016} treat the term ``particle-hole symmetry'' as synonymous with charge-conjugation symmetry, $C$, which is unitary! Given that the mathematical setting of the Tenfold Way is motivated primarily by physics applications to electrons in condensed matter, such an identification does not look convincing. In fact, charge conjugation $C$ transforms a universe with electrons and baryons into an anti-universe with positrons and anti-baryons while reversing the electromagnetic field, $(E,B) \mapsto (-E,-B)$. Thus it is a symmetry only of the trivial state of an empty universe, and it certainly is never a symmetry of any system of non-relativistic condensed matter! To patch up the discrepancy, one might consider identifying our $K$ with the $CT$ of the AZ classification. Alas, that would do injustice to $K$, which typically does not involve time reversal ($T: \; k \mapsto -k$) but rather a shift of the single-particle momenta (e.g., $k \mapsto k \pm \pi$) or some other transformation taken from a space group.
\end{itemize}

This is the first of a series of two articles. In the second article, the author will survey the Tenfold Way (a.k.a.\ the ``AZ classification'') built with the proper notion of particle-hole symmetry, as intended by the authors of the original work \cite{HHZ}.

\bigskip\noindent{\bf Acknowledgement.} The author acknowledges useful discussions with N. Read (2000), G.-M. Graf (2017), S. Kumar, A. Rosch, and J. Verbaarschot (2020).


\section*{References}
\begin{thebibliography}{11}
%
\bibitem{Affl-Hald} I.\ Affleck, F.D.M.\ Haldane, ``Critical theory of quantum spin chains'', {\it Phys.\ Rev.\ B} {\bf 36} (1987) 5291-5300
%
\bibitem{AKLT} I.\ Affleck, T.\ Kennedy, E.H.\ Lieb, H.\ Tasaki, ``Rigorous results on valence-bond ground states in antiferromagnets'', {\it Phys.\ Rev.\ Lett.} {\bf 59} (1987) 799-802
%
\bibitem{AffleckLieb} I.\ Affleck, E.H.\ Lieb, ``A proof of part of Haldane conjecture on spin chains'', {\it Lett.\ Math.\ Phys.} {\bf 12} (1986) 57-69
%
\bibitem{AMZ} A.\ Alldridge, C.\ Max, M.R.\ Zirnbauer, ``Bulk-boundary correspondence for disordered free-fermion topological phases'', {\it Commun.\ Math.\ Phys.}, DOI:10.1007/s00220-019-03581-7
%
\bibitem{AZ97} A.\ Altland, M.R.\ Zirnbauer, ``Non-standard symmetry classes in mesoscopic normal-/super\-con\-ducting hybrid systems'', {\it Phys.\ Rev.} B {\bf 55} (1997) 1142-1161
%
\bibitem{AR2007} F.\ Anfuso, A.\ Rosch, ``String order and adiabatic continuity of Haldane chains and band insula\-tors'', {\it Phys.\ Rev.\ B} {\bf 75} (2007) 144420
%
\bibitem{ASS-PRL} J.E.\ Avron, R.\ Seiler, B.\ Simon, ``Quantum Hall effect and relative index for projections'', {\it Phys.\ Rev.\ Lett.} {\bf 65} (1990) 2185-2188
%
\bibitem{BMF-2015} M.\ Barkeshli, M.\ Mulligan, M.P.A.\ Fisher, ``Particle-hole symmetry and the composite Fermi liquid'', {\it Phys.\ Rev.\ B} {\bf 92} (2015) 165125
%
\bibitem{BjorkenDrell} J.D.\ Bjorken, S.D.\ Drell, \emph{Relativistic  Quantum Mechanics} (McGraw-Hill, New York, 1964)
%
\bibitem{NNB} N.N.\ Bogoliubov, ``A new method in the theory of superconductivity'', {\it Sov.\ Phys.\ -- JETP} {\bf 7} (1958) 41-46
%
\bibitem{BottTu} R.\ Bott, L.\ Tu, \emph{Differential forms in algebraic topology} (Springer, New York, 1982)
%
\bibitem{CTSR2016} C.-K.\ Chiu, J.C.Y.\ Teo, A.P.\ Schnyder, S.\ Ryu, ``Classification of topological quantum matter with symmetries'', {\it Rev.\ Mod.\ Phys.} {\bf 88} (2016) 035005
%
\bibitem{deGennes} P.G.\ deGennes, {\it Superconductivity of Metals and Alloys} (W.A.\ Benjamin, New York, 1966)
%
\bibitem{dyson} F.J.\ Dyson, ``Threefold Way -- algebraic structure of symmetry groups and ensembles in quantum mechanics'', {\it J.\ Math.\ Phys.} {\bf 3} (1962) 1199-1215
%
\bibitem{FM2013} D.S.\ Freed, G.W.\ Moore, ``Twisted equivariant matter'', {\it Ann. Henri Poincar\'e} {\bf 14} (2013) 1927-2023
%
\bibitem{GZMMVM} S.D.\ Geraedts, M.P.\ Zaletel, R.S.K.\ Mong, M.A.\ Metlitski, A.\ Vishwanath, O.I.\ Motrunich, ``The half-filled Landau level: the case for Dirac composite fermions'', {\it Science} {\bf 352} (2016) 197-201
%
\bibitem{Girvin} S.M.\ Girvin, ``Particle-hole symmetry in the anomalous quantum Hall effect'', {\it Phys.\ Rev.\ B} {\bf 29} (1984) 6012-6014
%
\bibitem{Girvin-MacD} S.M.\ Girvin, A.M.\ MacDonald, ``Off-diagonal long range order, oblique confinement, and the fractional quantum Hall effect'', {\it Phys.\ Rev.\ Lett.} {\bf 58} (1987) 1252-1255
%
\bibitem{HaldaneConj} F.D.M.\ Haldane, ``Continuum dynamics of the 1D Heisenberg antiferromagnet: identification with the $\mathrm{O}(3)$ nonlinear sigma model'', {\it Phys.\ Lett.} {\bf 93}A (1983) 464-468
%
\bibitem{FDMH-Berry} F.D.M.\ Haldane, ``Berry curvature on the Fermi surface: anomalous Hall effect as a topological Fermi-liquid property'', {\it Phys.\ Rev.\ Lett.} {\bf 93} (2004) 206602
%
\bibitem{HLR} B.I.\ Halperin, P.A.\ Lee, N.\ Read, ``Theory of the half-filled Landau level'', {\it Phys.\ Rev.\ B} {\bf 47} (1993) 7312-7343
%
\bibitem{HHZ} P.\ Heinzner, A.\ Huckleberry, M.R.\ Zirnbauer, ``Symmetry classes of disordered fermions'',\\ {\it Commun.\ Math.\ Phys.} {\bf 257} (2005) 725-771
%
\bibitem{Jain} J.K.\ Jain, ``Composite-fermion approach for the fractional quantum Hall effect'', {\it Phys.\ Rev.\ Lett.} {\bf 63} (1989) 199-202
%
\bibitem{duality} A.\ Karch, D.\ Tong, ``Particle-vortex duality from 3D bosonization'', {\it Phys.\ Rev.\ X} {\bf 6} (2016) 031043
%
\bibitem{KZ-PS} R.\ Kennedy, M.R.\ Zirnbauer, ``Bott-Kitaev periodic table and the diagonal map'', {\it Physica Scripta} T{\bf 164} (2015) 014010
%
\bibitem{KZ} R.\ Kennedy, M.R.\ Zirnbauer, ``Bott periodicity of $\mathbb{Z}_2$ symmetric gapped free-fermion ground\\ states'', {\it Commun.\ Math.\ Phys.} {\bf 342} (2016) 909-963
%
\bibitem{Kitaev-BDI} A.Yu.\ Kitaev, ``Unpaired Majorana fermions in quantum wires'', {\it Phys.-Usp.} {\bf 44} (2001) 131-136
%
\bibitem{Laughlin} R.B.\ Laughlin, ``Anomalous quantum Hall effect: an incompressible quantum fluid with fractionally charged excitations'', {\it Phys.\ Rev.\ Lett.} {\bf 50} (1983) 1395-1398
%
\bibitem{LSM1961} E.\ Lieb, T.\ Schultz, D.\ Mattis, ``Two soluble models of an antiferromagnetic chain'' {\it Ann.\ Phys.} {\bf 16} (1961) 407-466
%
\bibitem{MFM2015} T.\ Morimoto, A.\ Furusaki, C.\ Mudry, ``Breakdown of the topological classification $\mathbb{Z}$ for gapped phases of noninteracting fermions by quartic interactions'', {\it Phys.\ Rev.\ B} {\bf 92} (2015) 125104
%
\bibitem{MP2015} S.\ Moudgalya, F.\ Pollmann, ``Fragility of symmetry protected topological phases on a Hubbard ladder'', {\it Phys.\ Rev.\ B} {\bf 91} (2015) 155128
%
\bibitem{MS-RMP} G.\ Murthy, R.\ Shankar, ``Hamiltonian theories of the fractional quantum Hall effect'', {\it Rev.\ Mod.\ Phys.} {\bf 75} (2003) 1101-1158
%
\bibitem{MS2015} G.\ Murthy, R.\ Shankar, ``$\nu = 1/2$ Landau level: half-empty versus half-full'', {\it Phys.\ Rev.\ B} {\bf 93} (2016) 085405
%
\bibitem{Nambu} Y.\ Nambu, ``Quasi-particles and gauge invariance in the theory of superconductivity'', {\it Phys.\ Rev.} {\bf 117} (1960) 648-663
%
\bibitem{NiemiSemenoff} A.J.\ Niemi, G.W.\ Semenoff, ``Axial-anomaly induced fermion fractionization and effective gauge-theory actions in odd-dimensional space-times'', {\it Phys.\ Rev.\ Lett.} {\bf 51} (1983) 2077-2080
%
\bibitem{PBTO} F.\ Pollmann, E.\ Berg, A.M.\ Turner, M.\ Oshikawa, ``Symmetry protection of topological order in one-dimensional quantum spin systems'', {\it Phys.\ Rev.\ B} {\bf 85} (2012) 075125
%
\bibitem{Prodan-Schuba} E.\ Prodan, H.\ Schulz-Baldes, {\it Bulk and boundary invariants for complex topological insulators}\\ (Mathematical Physics Studies, Springer, 2016)
%
\bibitem{QKS2016} R.\ Queiroz, E.\ Khalaf, A.\ Stern, ``Dimensional hierarchy of fermionic interacting topological phases {\it Phys.\ Rev.\ Lett.} {\bf 117} (2016) 206405
%
\bibitem{Read-CF1} N. Read, ``Theory of the half-filled Landau level'', {\it Semicond. Sci. Technol.} {\bf 9} (1994) 1859-1864
%
\bibitem{Read-CF2} N. Read, ``Recent progress in the theory of composite fermions near even-denominator filling factors'', {\it Surface Science} {\bf 361/362} (1996) 7-12
%
\bibitem{Read-nu1} N. Read, ``Lowest-Landau-level theory of the quantum Hall effect: the Fermi-liquid-like state for bosons at filling factor one'', {\it Phys.\ Rev.\ B} {\bf 58} (1998) 16262-16290
%
\bibitem{Redlich} A.N.\ Redlich, ``Gauge noninvariance and parity nonconservation of three-dimensional fermions'', {\it Phys.\ Rev.\ Lett.} {\bf 52} (1984) 18-21
%
\bibitem{SRFL2010} S.\ Ryu, A.P.\ Schnyder, A.\ Furusaki, A.W.W.\ Ludwig, ``Topological insulators and superconductors: tenfold way and dimensional hierarchy'', {\it New.\ J.\ Phys.} {\bf 12} (2010) 065010
%
\bibitem{duality4} N.\ Seiberg, T.\ Senthil, C.\ Wang, E.\ Witten, ``A duality web in $2+1$ dimenions and condensed matter physics, {\it Ann.\ Phys.} {\bf 374} (2016) 395-433
%
\bibitem{DTSon} D.T.\ Son, ``Is the composite fermion a Dirac particle?'',  {\it Phys.\ Rev.\ X} {\bf 5} (2015) 031027
%
\bibitem{SSH} W.P.\ Su, J.R.\ Schrieffer, A.J.\ Heeger, ``Solitons in polyazetylene'', {\it Phys.\ Rev.\ Lett.} {\bf 42} (1979)\\ 1698-1701
%
\bibitem{Thaller} B.\ Thaller, {\it The Dirac equation} (Springer, Berlin, Heidelberg, 1991)
%
\bibitem{vanEnk} S.J.\ van Enk, ``Angular momentum in the quantum Hall effect and its sign'', arXiv:1906.00342
%
\bibitem{Ver94} J.\ Verbaarschot, ``Spectrum of the QCD Dirac operator and chiral random-matrix theory'', {\it Phys.\ Rev.\ Lett.} {\bf 72} (1994) 2531-2533
%
\bibitem{VMP2017} R.\ Verresen, R.\ Moessner, F.\ Pollmann, ``One-dimensional symmetry protected topological phases and their transitions'', {\it Phys.\ Rev.\ B} {\bf 96} (2017) 165124
%
\bibitem{DCF-WS} C.\ Wang, T.\ Senthil, ``Half-filled Landau level, topological insulator surfaces, and three-dimensional quantum spin liquids'', {\it Phys.\ Rev.\ B} {\bf 93} (2016) 085110
%
\bibitem{WS-AIII} C.\ Wang, T.\ Senthil, ``Interacting fermionic topological insulators / superconductors in three\\ dimensions'', {\it Phys.\ Rev.\ B} {\bf 89} (2014) 195124
%
\bibitem{WCHS-PRX} C.\ Wang, N.R.\ Cooper, B.I.\ Halperin, A.\ Stern, ``Particle-hole symmetry in the Fermion-Chern-Simons and Dirac descriptions of a half-filled Landau level'', {\it Phys.\ Rev.\ X} {\bf 7} (2017) 031029
%
\bibitem{Weinberg} S.\ Weinberg, ``The Quantum Theory of Fields'', vol.\ 1 (Cambridge University Press, 1995)
%
\bibitem{OHRMT} M.R.\ Zirnbauer, ``Symmetry classes'' in {\it The Oxford Handbook of Random-Matrix Theory} (Oxford University Press, 2011); arXiv:1001.0722
%
\bibitem{MRZ-2016} M.R.\ Zirnbauer, ``On symmetry-protected topological states: from free fermions to the Haldane phase'', {\it Talk given in Oxford 11/2016, Z\"urich 12/2016, Stony Brook 01/2017, Hamburg 06/2017}
%
\bibitem{MRZ-LF} M.R.\ Zirnbauer, ``The Tenfold Way of disordered fermions as a Lingua Franca for band structures and topological insulators'', \emph{in preparation}
%
\bibitem{footnote0} We insist on the basic principle that (closed) quantum systems must have excitation energies which are all positive relative to a stable ground state. Any ``symmetry'' operation that anti-commutes with the true Hamiltonian $H$ would imply negative excitation energies, hence a contradiction.
%
\bibitem{footnote} The letter $K$ stands for ``Konjugation'', the German equivalent of the English word ``conjugation''. I very much hope that I got it right this time and the community will catch on. (One objection might be that physicists often use $K$ to denote complex conjugation. In my defense, I would say that complex conjugation of a Hermitian vector space is not invariantly defined but changes under unitary transformations of the basis; thus it is not an object deserving of much respect.)
%
\bibitem{footnote-DIS} Note that we do not have a mathematical proof (in fact, the gapped nature of the antiferromagnetic spin-1 Heisenberg chain has not been proved rigorously), just a very plausible physics argument.
%
\bibitem{footnote-OR} On any orientable space-time manifold $\mathcal{M}$, the passage from $a_{\rm Son}$ to $a_{\rm HLR}$ would involve the step of tensoring with a section of the orientation line bundle over $\mathcal{M}$.
%
\bibitem{Poincare} The underlying theorem is that the Chern class of a complex line bundle (or equivalently, the Euler class of a rank-2 real vector bundle) over an oriented 2-manifold $M$ is Poincar\'e-dual to the zero locus of a transversal section \cite{BottTu}. This translates to the statement that the number of our wave function zeroes, counted with algebraic multiplicity, equals the total magnetic flux through $M$ in units of the flux quantum (a.k.a.\ Chern number).
%
\bibitem{foot:anomal} If the charge total is half-integer, $Q / q \in  \mathbb{Z} + 1/2$, the partition function for the $2+1$ dimensional Dirac fermion (\ref{eq:S-dual}) fails to be invariant under large gauge transformations -- one says that it suffers from the parity anomaly \cite{Redlich,NiemiSemenoff}. Such an anomaly is physically acceptable for the surface theory of a topological insulator or for a UV-incomplete theory such as that of the lowest Landau level. The parity anomaly is avoided by restricting the allowed charges to $Q / q \in \mathbb{Z}$.
\end{thebibliography}
\end{document}